\documentclass[prb,twocolumn,twoside,letterpaper,showpacs,superscriptaddress,nobibnotes,byrevtex]{revtex4}

\usepackage{graphicx,color} 

\usepackage[mathlf,minionint,openg]{MinionPro} 
\usepackage{eucal}

\usepackage{setspace} 
\usepackage{dcolumn}

\newcommand{\acro}[1]{\textsc{\MakeLowercase{#1}}}

\newcommand{\me}{\mathrm{e\,}}
\newcommand{\mi}{\mathrm{i}}

\newcommand{\dif}{\,\mathrm{d}}

\renewcommand{\vec}[1]{\mathbf{#1}}

\newcommand{\abs}[1]{\lvert#1\rvert}
\newcommand{\norm}[1]{\lVert#1\rVert}

\newcommand{\braket}[2]{\langle #1 | #2\rangle}
\newcommand{\braketw}[3]{\langle\!\langle #1 | #2\rangle\!\rangle_{#3}}
\newcommand{\pseudo}[2]{\left[ #1 \big | #2 \right]}

\newcommand{\op}[1]{\mathsf{#1}}
\newcommand{\field}[3]{#1_{#2}^{(#3)}}

\DeclareMathOperator{\Real}{Re}
\DeclareMathOperator{\Imag}{Im}

\definecolor{orange}{rgb}{1.0,0.55,0.0}

\newcommand{\rb}[1]{\raisebox{1.5ex}[0pt]{#1}}

\begin{document}
\title{Modal Analysis and Coupling in\\ Metal-Insulator-Metal Waveguides}

\author{\c{S}{\"u}kr{\"u}~Ekin~Kocaba\c{s}}
\email{kocabas@ieee.org}
\affiliation{Ginzton Laboratory, Stanford University, Stanford, CA 94305}

\author{Georgios~Veronis}
\email{gveronis@lsu.edu}
\affiliation{Department of Electrical and Computer Engineering and Center for Computation and Technology, Louisiana State University, Baton Rouge, LA 70803}

\author{David~A.B.~Miller}
\email{dabm@ee.stanford.edu}
\affiliation{Ginzton Laboratory, Stanford University, Stanford, CA 94305}

\author{Shanhui~Fan}
\email{shanhui@stanford.edu}
\affiliation{Ginzton Laboratory, Stanford University, Stanford, CA 94305}

\begin{abstract}
This paper shows how to analyze plasmonic metal-insulator-metal waveguides using the full modal structure of these guides. The analysis applies to all frequencies, particularly including the near infrared and visible spectrum, and to a wide range of sizes, including nanometallic structures. We use the approach here specifically to analyze waveguide junctions. We show that the full modal structure of the metal-insulator-metal (\acro{MIM}) waveguides---which consists of real and complex discrete eigenvalue spectra, as well as the continuous spectrum---forms a complete basis set. We provide the derivation of these modes using the techniques developed for Sturm-Liouville and generalized eigenvalue equations. We demonstrate the need to include all parts of the spectrum to have a complete set of basis vectors to describe scattering within \acro{MIM} waveguides with the mode-matching technique. We numerically compare the mode-matching formulation with finite-difference frequency-domain analysis and find very good agreement between the two for modal scattering at symmetric \acro{MIM} waveguide junctions. We touch upon the similarities between the underlying mathematical structure of the \acro{MIM} waveguide and the $\mathcal{PT}$ symmetric quantum mechanical pseudo-Hermitian Hamiltonians. The rich set of modes that the \acro{MIM} waveguide supports forms a canonical example against which other more complicated geometries  can be compared. Our work here encompasses the microwave results, but extends also to waveguides with real metals even at infrared and optical frequencies.
\end{abstract}




\maketitle
 
\section{Introduction}
Waveguides have long been used to controllably direct energy flow between different points in space. Understanding the way waves propagate in waveguides led to a multitude of creative designs---all the way from the pipe organ to light switches used in fiber optic communications. In optics, recently there has been a growing interest in making use of the dielectric properties of metals to guide electromagnetic energy by using sub-wavelength sized designs that work in the infrared and the visible bands of the spectrum. One of the motivations for doing photonic research using metals is to find the means to integrate electronic devices with sizes of tens of nanometers with the relatively much larger optical components---so that some of the electrons used in the communication channels between electrical circuitry can be replaced by photons for faster and cooler operation \cite{Ozbay2006}.

Whereas the use of metals for directing electromagnetic energy is relatively new in optics, sub-wavelength guiding of light by metals is the norm in the microwave domain. Even though the permittivity of metals can  be large in magnitude at both microwave and optical frequencies, the characteristics of the permittivity are quite different in the two frequency regimes.

In the microwave regime, electrons go through multiple collisions with the ions of the lattice during an electromagnetic cycle according to the phenomenological Drude model of electrons. Therefore, the electron movement is a drift motion where the \textsl{velocity} of electrons is proportional to the applied field strength (Ref. \onlinecite{Ashcroft1976}, Chp. 1). As a result, the induced dipole moment density and hence the permittivity is a large, negative\cite{footnote_1} imaginary number. On the other hand, at optical frequencies below the plasma oscillation frequency, electrons go through a negligible number of collisions during an electromagnetic cycle and this time \textsl{acceleration} of electrons is proportional to the applied field strength which then results in a permittivity that can be substantially a large real negative number. Above the plasma frequency, the induced dipole moment density is very low and the permittivity is predominantly a positive real number less than one (Ref. \onlinecite{Bohren1983}, Chp. 9).

The dielectric slab and the parallel plate (i.e. consisting of two parallel perfectly conducting metal plates) waveguides are the two canonical examples of waveguiding theory. If we have a layered \textsl{metal-insulator-metal} \acro{(MIM)} geometry, it is possible to smoothly transition from the dielectric slab to the parallel plate waveguide by reducing the frequency of operation, and therefore varying the metallic permittivity $\epsilon_m$.

\begin{figure}[htb]
\centering
\includegraphics[keepaspectratio=true,width=8.6cm]{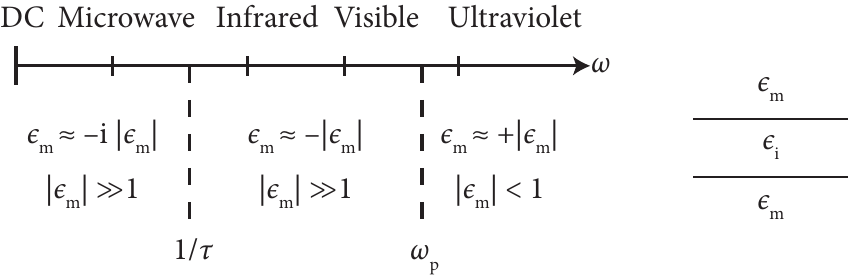} 
\caption{The change in the metallic permittivity $(\epsilon_m)$ as the frequency of operation $(\omega)$ is varied leads to an evolution from the parallel plate waveguide at low frequencies to the dielectric slab waveguide at high frequencies for the two dimensional metal-insulator-metal geometry. We are assuming that $\epsilon_m$ is a Drude model metal where $\tau$ is the average time between collisions among the ions of the lattice and the free electrons, $\omega_p$ is the plasma oscillation frequency.}\label{figOverview}
\end{figure}

At frequencies above the plasma frequency, the metal has a permittivity $\epsilon_m<1$ whereas the insulator has $\epsilon_i \ge 1$. We illustrate the geometry in the inset of Fig. \ref{figOverview}.

The physical modes that the dielectric slab waveguide supports fall into two sets: guided modes and radiation modes (Ref. \onlinecite{Marcuse1991}, Chp. 1). Guided modes consist of a countable, finite set, i.e. there is only a finite number of discrete guided modes. Radiation modes consist of a non-countable, infinite set, i.e. they form a continuum. The combination of these two sets of modes form a complete and orthogonal basis set.

Now suppose that we change our operation frequency to one which is very close to the \acro{DC} limit where $\epsilon_m$ is an arbitrarily large, negative, imaginary number. In this limit, we can approximate the metal as a \textsl{perfect electric conductor} (\acro{PEC}) where $|\epsilon_m|\rightarrow\infty$. Such an approximation then gives us the parallel plate waveguide of the microwave domain. Unlike the dielectric slab, the parallel plate geometry is bounded in the transverse dimension---fields are not allowed to penetrate into the \acro{PEC}. There are infinitely many discrete modes of the parallel plate, all of which have sinusoidal shapes, and there are no continuum modes. The collection of the infinitely many discrete modes forms a complete orthogonal basis set.

In this paper we will investigate the modal structure of the two dimensional \acro{MIM} waveguide in the infrared regime where $\epsilon_m$ is primarily a large, negative real number. The geometry of the \acro{MIM} waveguide is exactly the same as the one in the inset of Fig. \ref{figOverview}. The only difference between the parallel plate, \acro{MIM} and the dielectric slab waveguides is in the numerical value of $\epsilon_m$, which depends on the frequency of operation.

There have been numerous studies of the \acro{MIM} waveguide in the literature\cite{Davis1965,Davis1966,Economou1969,Takano1972,Kaminow1974,Prade1991,Villa2001,Qing2005,Dionne2006,Ginzburg2006,Kim2006a,Gordon2006,Feigenbaum2007a,Kurokawa2007,Wang2007,Sun2007,Zakharian2007,Sturman2007}. The fact that light can be guided within a deep subwavelength volume over a very wide range of wavelengths is one of the primary reasons why the \acro{MIM} geometry has attracted so much attention. The full set of modes that the \acro{MIM} waveguide supports---real and complex discrete modes as well as a continuous set of modes---has only very recently been published\cite{Sturman2007}. For other geometries, it has been shown that, in general, waveguides support real, complex and continuous sets of modes\cite{Tamir1963,Paiva1991,Jablonski1994,Shu2006,Kim2006}. In this work, we will provide the detailed mathematical framework to analyze the modal structure of the \acro{MIM} waveguide and emphasize how it is a hybrid between the parallel plate and the dielectric slab waveguides.

Operator theory will be the basis of the mathematical tool set with which we will analyze \acro{MIM} waveguides. In Section \acro{II}, we will introduce the notation and make some definitions pertaining to the operators in infinite dimensional spaces. In Section \acro{III} we will derive the discrete and continuum modes supported by the \acro{MIM} waveguide and show that the underlying operators are pseudo-Hermitian. In Section \acro{IV} we will demonstrate that the modes we report form a complete basis set via example calculations using the mode-matching technique. In Section \acro{V} we will discuss our results and underline some of the relevant developments in mathematics from both quantum mechanics and microwave theory with the hope of expanding our tools of analysis. Lastly, in Section \acro{VI} we will draw our conclusions.

\section{Some Definitions}
Throughout the paper, we will be using nomenclature from operator theory. In this section, we will define the terminology and introduce the notation which will be used in the following sections\cite{footnote_2}\nocite{Chew1990}. The reader well versed in operator theory can directly skip to the next section.

A \textsl{linear vector space} is a space which is closed under the operations of addition and of multiplication by a scalar. We will call the elements of the space \textsl{vectors}. Spaces need not be finite dimensional---infinite dimensional vector spaces are also possible. For instance, the collection of all square integrable functions $f(x)$ defined on an interval $a<x<b$ forms an infinite dimensional vector space.

The \textsl{inner product} is a scalar valued function of two vectors $f$ and $g$, written $\braket{f}{g}$ with the following properties
\begin{align*}
	\begin{split}
	\braket{f}{g}={}&\braket{g}{f}^*\\
	\braket{\alpha_1 f + \alpha_2 g}{h} ={}&\alpha_1^*\braket{f}{h} + \alpha_2^*\braket{g}{h}\\
	\braket{f}{f} > {}& 0 \qquad \text{if $f\neq 0$.}
	\end{split} 
\end{align*}
Here $(\cdot)^*$ denotes complex conjugation, $\alpha_{\{1,2\}}$ are arbitrary complex numbers and $f$, $g$, $h$ denote arbitrary members of the linear vector space $\mathcal{S}$. 

For the infinite dimensional vector space of square integrable functions one possible definition of the inner product is
\begin{align}
	\braket{f}{g}=\int_a^b f^*(x) g(x) \dif x.\label{eq2}
\end{align}
A linear vector space with an inner product is called an \textsl{inner product space}. In such spaces the \textsl{norm of a vector} $f$ is defined as
\begin{align}
\norm{f} = \sqrt{\braket{f}{f}}. \label{eq3}
\end{align}
This is also known as the $\mathcal{L}^2$ norm, to denote square integrability in the sense of Lebesgue. By using the norm of a vector space, we can define the distance between its vectors $f$ and $g$ as $d(f,g)=\norm{f-g}$ which is always nonzero if $f \neq g$. Here, $d(f,g)$ is called the \textsl{metric}---the measure of distance between vectors---of the inner product space. Suppose that $\mathcal{F}$ and $\mathcal{G}$ are two subsets of the inner product space $\mathcal{S}$ and that $\mathcal{F}$ is also a subset of $\mathcal{G}$, i.e. $\mathcal{F} \subset \mathcal{G} \subset \mathcal{S}$. $\mathcal{F}$ is said to be \textsl{dense} in $\mathcal{G}$, if for each $g\in\mathcal{G}$ and $\varepsilon>0$, there exists an element $f\in\mathcal{F}$ where $d(f,g)<\varepsilon$ (Ref. \onlinecite{Hanson2002}, pp. 94--95).

A vector space $\mathcal{S}$ is \textsl{complete} if all converging sequences of vectors $f_n(x)$ converge to an element $f \in \mathcal{S}$. An inner product space which is complete when using the norm defined by \eqref{eq2}-\eqref{eq3} is called a \textsl{Hilbert space}.

An \textsl{operator} $\op{L}$ is a mapping that assigns to a vector $f$ in a linear vector space $\mathcal{S}_1$ another vector in a different vector space $\mathcal{S}_2$ which we denote by $\op{L}f$ (most often $\mathcal{S}_1=\mathcal{S}_2$). An operator is \textsl{linear} if $\op{L}(\alpha_1 f + \alpha_2 g) = \alpha_1 \op{L} f + \alpha_2 \op{L} g$ for arbitrary scalars $\alpha_{\{1,2\}}$ and vectors $f$, $g$. The \textsl{domain} of an operator $\op{L}$ is the set of vectors $f$ for which the mapping $\op{L}f$ is defined. The \textsl{range} of an operator $\op{L}$ is the set of vectors $g=\op{L}f$ for all possible values of $f$ in the domain of $\op{L}$.  A linear operator is \textsl{bounded} if its domain is the entire linear space $\mathcal{S}$ of vectors $f$ and if there exists a single constant $C$ such that $\norm{\op{L}f}< C\norm{f}$. Otherwise the operator is \textsl{unbounded}. The differential operator is a classical example of an unbounded operator (Ref. \onlinecite{Kreyszig1978}, pp. 93--94). $\op{L}$ is \textsl{positive (negative) definite} if $\braket{f}{\op{L}f}>0$ $\bigl(\braket{f}{\op{L}f}<0\bigr)$ for all possible $f$. Otherwise, $\op{L}$ is \textsl{indefinite}.

A linear bounded operator $\op{L}^\dag$ is said to be the \textsl{adjoint} of $\op{L}$ if, for all $f$ and $g$ in $\mathcal{S}$ the condition $\braket{g}{\op{L}f} = \braket{\op{L}^\dag g}{f}$ is satisfied. If $\op{L}=\op{L}^\dag$ then $\op{L}$ is said to be \textsl{self-adjoint}. If the operator $\op{L}$ is unbounded, then the equality $\braket{g}{\op{L}f} = \braket{\op{L^\dag} g}{f}$ defines a \textsl{formal} self-adjoint (Ref. \onlinecite{Hanson2002}, Sec. 3.4.1).

Suppose we have a set of orthonormal vectors $\{f_n\}$ which span the Hilbert space $\mathcal{H}$. Then, we can expand any vector $g \in \mathcal{H}$ as $g = \sum_n \braket{f_n}{g} f_n$. Similarly, any linear bounded operator $\op{L}$ acting on $g$ results in
\begin{align*}
	\op{L}g = {}& \sum_n \braket{f_n}{g} \op{L} f_n = \sum_{n,m} \braket{f_n}{g} \braket{f_m}{\op{L} f_n} f_m\\
					= {}& \sum_{m,n} f_m \braket{f_m}{\op{L} f_n} \braket{f_n}{g} 	
\end{align*}
where we expanded $\op{L} f_n$ in terms of the basis $\{f_m\}$ to get to the last line. Once we choose a complete orthonormal basis set, we can describe the action of $\op{L}$ on any vector $g$ by the product of an infinite dimensional matrix with elements $\braket{f_m}{\op{L} f_n}$ and an infinite dimensional vector with elements $\braket{f_n}{g}$---a generalization of regular matrix multiplication. The infinite dimensional matrix is called the \textsl{representation} of $\op{L}$ in $\{f_n\}$. If the matrix for $\op{L}$ is diagonal, then we call that the \textsl{spectral representation} (Ref. \onlinecite{Friedman1990}, p. 110).

\begin{figure}[htb]
\centering
\includegraphics[keepaspectratio=true,width=8.6cm]{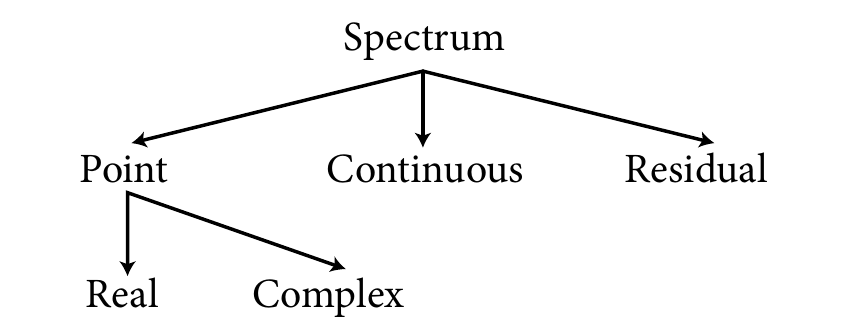}
\caption{Classification of the spectrum for an operator in a general space. Point spectrum is sometimes called the discrete spectrum.}\label{figSpectrum}
\end{figure}

The spectral representation for an operator $\op{L}$ depends on the study of the inverse of the operator $\op{L}-\lambda$, which we will denote by $(\op{L}-\lambda)^{-1}$, for all complex values of $\lambda$ (Ref. \onlinecite{Friedman1990}, p. 125). Let the domain and range of $\op{L}$ be denoted by $\mathcal{D}_L$ and $\mathcal{R}_L$. The \textsl{point (discrete) spectrum} is the set of $\lambda$ for which $(\op{L}-\lambda)^{-1}$ does not exist. The \textsl{continuous spectrum} is the collection of $\lambda$ for which $(\op{L}-\lambda)^{-1}$ exists and is defined on a set dense in $\mathcal{R}_L$, but for which it is unbounded. The \textsl{residual spectrum} is the collection of $\lambda$ for which $(\op{L}-\lambda)^{-1}$ exists (it may or may not be bounded), but for which it is not defined on a set dense in $\mathcal{R}_L$. The \textsl{spectrum} of $\op{L}$ consists of values of $\lambda$ which belong to either the point, continuous or the residual spectrum (Ref. \onlinecite{Kreyszig1978}, p. 371 and Ref. \onlinecite{Locker2000}, p. 21). We summarized the taxonomy of the spectrum in Fig. \ref{figSpectrum}.

\section{Spectrum}
\begin{figure}[b]
\centering
\includegraphics[keepaspectratio=true,width=8.6cm]{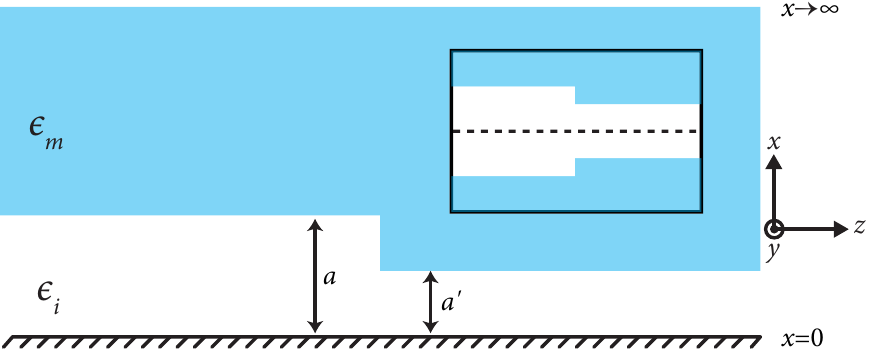}
\caption{(Color online) Geometry for the even modes of the \acro{MIM} waveguide. The $x=0$ plane contains a fictitious perfect electric conductor to simplify the problem when dealing only with even \acro{TM} modes of the guide. This fictitious \acro{MIM} waveguide is equivalent to an actual guide with an insulator thickness of $2a$. The inset shows the equivalent symmetric junction of two \acro{MIM} waveguides. The dashed line in the inset is the plane of symmetry, which is where the fictitious \acro{PEC} layer is introduced.}\label{figGeometry}
\end{figure}

After having defined the necessary terminology, in this section we will derive the modal structure (spectrum) of the \acro{MIM} waveguide. We will specifically focus on the even modes of the waveguide, for which the \textsl{transverse magnetic} (\acro{TM}) field component is an even function of the transverse coordinate, $x$. The reason why we focus on even modes is that we will be analyzing the scattering of the main, even mode of the \acro{MIM} waveguide---which is also a \acro{TM} mode---off of a symmetric junction with a different sized \acro{MIM} waveguide. Due to the symmetry of the problem at hand, even modes will be sufficient. We could also solve for the case of the odd modes by a similar approach, but we omit that explicit solution for reasons of space. Evenness of the function is achieved by putting a fictitious \textsl{perfect electric conductor} (\acro{PEC}) at the $x=0$ plane of the waveguide, which forces the tangential electric field $E_z$ to be an odd function, and the magnetic field $H_y$ to be an even function of $x$. In other words, the modes of this fictitious waveguide with the \acro{PEC} at $x = 0$ are mathematically the same as the even modes of the actual waveguide of interest, and so we will work with this hypothetical waveguide for our mathematics. The geometry is as shown in Fig. \ref{figGeometry}. $\epsilon_m$ refers to the permittivity of the metal region and $\epsilon_i$ of the insulator region. At infrared frequencies, $\epsilon_m$ is a complex number with a large, negative real part and a relatively small imaginary part (the sign of which is determined by the time convention used, being negative for an $\exp(\mi \omega t)$ time dependence).

Let us begin with Maxwell's equations for fields that have an $\exp(\mi \omega t)$ time dependence. 
\begin{equation}
\begin{split}
\begin{aligned}
\nabla \times \vec{E}(\vec{r})={}&  -&\mspace{-18mu}&\mi \omega \mu(x) 			\vec{H}(\vec{r})\\
\nabla \times \vec{H}(\vec{r})={}& {}&\mspace{-18mu}&\mi \omega \epsilon(x) \vec{E}(\vec{r}).
\end{aligned}
\end{split}\label{eq5}
\end{equation}
The \acro{MIM} waveguide is a two dimensional structure which does not have any variation in the $y$ direction. Therefore, we can eliminate all the derivatives with respect to $y$ in Maxwell's equations. Furthermore, our study will be based on the \acro{TM} modes which only have the $H_y$, $E_x$ and $E_z$ field components. Also, the uniformity of the waveguide in the $z$ direction leads to $\exp(-\mi k_z z)$ as the space dependence in $z$ by using the separation of variables technique for differential equations ($k_z$ may, however, be a complex number). After simplifying the curl equations in \eqref{eq5}, we have the following relationships between the different field components
\begin{align}
\begin{split}
	\mi \omega \mu(x) H_y(x) =& \mi k_z E_x(x) + \frac{\dif}{\dif x} E_z(x)\\
	\mi k_z H_y(x) =& \mi \omega \epsilon(x) E_x(x) \\
	\frac{\dif}{\dif x}H_y(x) =& \mi \omega \epsilon(x) E_z(x).
\end{split} \label{eq6}
\end{align}
Using these equations we get the following differential equation for $H_y$
\begin{align}
	\left(\epsilon(x) \frac{\dif}{\dif x} \frac{1}{\epsilon(x)} \frac{\dif}{\dif x} 
		+ \omega^2 \mu(x) \epsilon(x) \right )H_y = k_z^2  H_y \label{eq7}
\end{align}
and since $E_z(0)=0$ by the \acro{PEC} wall at $x=0$, the boundary condition for $H_y$ under \eqref{eq6} becomes $\left. \frac{\dif}{\dif x}H_y(x) \right|_{x=0} = 0$. The equation \eqref{eq7} is in the \textsl{Sturm-Liouville} form (Ref. \onlinecite{Hanson2002}, ch. 5).

\subsection{Point Spectrum}
The standard approach (Ref. \onlinecite{Hanson2002}, Chap. 5) in the calculation of the point spectrum of a Sturm-Liouville equation as in \eqref{eq7} starts with a redefinition of the space to $\mathcal{L}^2_{\epsilon}$---the set of all \textsl{weighted} Lebesgue square-integrable functions such that
\begin{align*}
	\int \abs{f(x)}^2\frac{1}{\epsilon(x)} \dif x < \infty
\end{align*}
which implies that the boundary condition at infinity should be $\lim_{x \rightarrow \infty} H_y(x) = 0$. The inner-product in $\mathcal{L}^2_\epsilon$, denoted by $\braketw{\cdot}{\cdot}{\epsilon}$, is then defined as
\begin{align}
	\braketw{f}{g}{\epsilon}=\int f^*(x)g(x) \frac{1}{\epsilon(x)}\dif x. \label{eq8}
\end{align}
In order to have a \textsl{definite} metric for $\mathcal{L}^2_\epsilon$, the inner product should be such that $\braketw{f}{f}{\epsilon}>0$ for all $f\neq0$ so that the norm of any non-zero vector will be a positive quantity. This in turn implies that to have a definite metric, $\epsilon(x)$ should be a real and positive number for all $x$. Within the Hilbert space obtained by our choice of the inner product $\braketw{\cdot}{\cdot}{\epsilon}$, we can write \eqref{eq7} as $\op{L}H_y=k_z^2 H_y$. The operator
\begin{align*}
	\op{L} = \epsilon(x) \frac{\dif}{\dif x} \frac{1}{\epsilon(x)} \frac{\dif}{\dif x} + \omega^2 \mu(x) \epsilon(x)
\end{align*}
is self-adjoint since $\braketw{f}{\op{L}g}{\epsilon} = \braketw{\op{L}f}{g}{\epsilon}$ for all $f$ and $g$ as long as $\epsilon(x)\in\mathbb{R}$ and $\epsilon(x)>0$. One can then easily prove that the point spectrum of $\op{L}$ is purely real (Ref. \onlinecite{Churchill1941}, p. 50). The lossless dielectric slab waveguide, which satisfies all the criteria we mentioned, therefore has a purely real point spectrum.

Unfortunately, the arguments above fail for the \acro{MIM} waveguide system since the condition $\epsilon(x)>0$ is no longer satisfied\cite{Sturman2007}. The dielectric constants of metals can have negative real parts at some frequencies (e.g., in the infrared and visible regions), and generally also have imaginary components corresponding to loss, especially at optical frequencies.  We will now separately analyze the lossy and the lossless metal cases.

\subsubsection{Lossless Case}
Since $\abs{\Imag(\epsilon_m)} \ll \abs{\Real(\epsilon_m)}$ at infrared frequencies, it is worthwhile investigating the case of real, negative permittivity, i.e.,  $\epsilon_m=-\abs{\epsilon_m}$. The standard Sturm-Liouville theory is not applicable in this case, because it requires the weighting function $\epsilon(x)$ to have the same sign over its entire domain of definition (Ref. \onlinecite{Churchill1941}, p. 50). However, $\epsilon_i > 0$ whereas $\epsilon_m < 0$ for the \acro{MIM} waveguide, under the approximation of negligible loss. The definition of the inner-product \eqref{eq8} becomes \textsl{indefinite} in this case, since we can have $\braketw{f}{f}{\epsilon}\le0$ for some $f\neq 0$. As a result, we no longer can operate in the Hilbert space. The space of functions with an indefinite metric is called the \textsl{Krein space}. In contrast to the Hilbert space case, the spectrum of the self-adjoint operators in Krein spaces is, in general, not real (Ref. \onlinecite{Zettl2005}, p. 220). An early analysis of a real Sturm-Liouville equation with a complex point spectrum can be found in Ref. \onlinecite{Richardson1918}.

To prove that \eqref{eq7} accepts complex solutions even when $\epsilon(x)\in\mathbb{R}$, let us work in the well defined $\mathcal{L}^2$ space with an inner-product as defined in \eqref{eq2}. Because $\braket{\cdot}{\cdot}$ is always definite, we are back in the Hilbert space, but $\op{L}$ is no longer self-adjoint in $\mathcal{L}^2$. Two integrations by parts\cite{footnote_3} give $\op{L}^\dag$ as
\begin{align*}
	\op{L}^\dag = \frac{\dif}{\dif x} \frac{1}{\epsilon(x)} \frac{\dif}{\dif x} \epsilon(x) 
		+ \omega^2 \mu(x) \epsilon(x)
\end{align*}
with boundary conditions
\begin{equation*}
\begin{aligned}
	H_y(x) &\Bigr|_{x \rightarrow \infty} \mspace{-18mu}&&= 0\\
	\frac{\dif}{\dif x}\epsilon(x) H_y(x) &\Bigr|_{x=0} \mspace{-18mu}&&= 0.
\end{aligned}
\end{equation*}
We see that $\op{L}^\dag = \epsilon^{-1} \op{L} \epsilon$ which makes $\op{L}$ by definition \textsl{pseudo-Hermitian} \cite{Mostafazadeh2002b,Mostafazadeh2008}. It has been proved that a pseudo-Hermitian operator does not have a real spectrum if $\mathcal{L}^2_\epsilon$ is indefinite (Ref. \onlinecite{Mostafazadeh2002b}, Th. 3).

Alternatively, we can approach the problem by defining
\begin{align*}
	\op{L}^\prime = \frac{\dif}{\dif x} \frac{1}{\epsilon(x)} \frac{\dif}{\dif x}
		+ \omega^2 \mu(x)
\end{align*}
and rewriting \eqref{eq7} as
\begin{align*}
	\op{L}^\prime H_y = \frac{k_z^2}{\op{\epsilon}} H_y.
\end{align*}
Using \eqref{eq2} it can be shown that $\op{L}^\prime$ \textsl{is} self-adjoint in $\mathcal{L}^2$ so that 
\begin{align*}
	\braket{H_{y2}}{\op{L}^\prime H_{y1}} = \braket{\op{L}^\prime H_{y2}}{ H_{y1}}
\end{align*}
which leads to the \textsl{generalized eigenvalue problem} for the self-adjoint operators $\op{L}^\prime$ and $\op{\epsilon}^{-1}$ as
\begin{align*}
	\op{L}^\prime H_y = k_z^2 \op{\epsilon}^{-1} H_y
\end{align*}
where $k_z^2$ is the eigenvalue. The point spectrum of the self-adjoint generalized eigenvalue problem will be complex only if \textsl{both} $\op{L}^\prime$ and $\op{\epsilon}^{-1}$ are indefinite (Ref. \onlinecite{Mrozowski1997}, p. 38). The indefiniteness of $\op{\epsilon}^{-1}$ is trivial because epsilon can be a positive or negative quantity now. To show that $\op{L}^\prime$ is indefinite, observe that by using the boundary conditions in integration by parts, one can get
\begin{align*}
	\braket{H_{y}}{\op{L}^\prime H_{y}} = \int_0^\infty \left( \omega^2 \mu \abs{H_y(x)}^2
		- \frac{1}{\epsilon(x)} \left|\frac{\dif H_y(x)}{\dif x}\right|^2 \right) \dif x
\end{align*}
which can be positive or negative depending on the choice of $H_y(x) \in \mathcal{L}^2$. Therefore, $\op{L}^\prime$ is indefinite, and $k_z^2$ will accept complex values. Note that the classification of the point spectrum into the real and the complex categories is based on $k_z^2$ and not $k_z$. Hence, the set of modes with purely real and negative $k_z^2$---which leads to a purely imaginary $k_z$---are categorized as real modes in this approach.

\subsubsection{Lossy Case}
As we mentioned earlier, $\epsilon_m$ has an imaginary part. As a result, for those cases when neglecting the imaginary part of $\epsilon_m$ is not desired, $\op{L}$ cannot be made self-adjoint by a redefinition of the inner-product. Therefore, the point spectrum---the set of $k_z^2$ for which $(\op{L}-k_z^2)$ does not have an inverse---will be complex. A general classification of the spectrum for non-self-adjoint operators is still an open problem (Ref. \onlinecite{Zettl2005}, p. 301 and Ref. \onlinecite{Davies2002}). Also, completeness of the spectrum is difficult to prove. However, the \acro{MIM} waveguiding problem can be shown to have a spectrum which forms a complete basis set even when $\op{L}$ is non-self-adjoint (Ref. \onlinecite{Hanson2002}, pp. 333-334, Th. 5.3).

\subsubsection{Mode Shape}
The dispersion equation that should be solved in order to find the $k_z$ values for the modes of the \acro{MIM} waveguide is derived by satisfying the continuity of tangential electric and magnetic fields at material boundaries and applying the boundary conditions. We refer the reader to Ref. \onlinecite{Hanson2002}, pp. 462--470 and Refs. \onlinecite{Economou1969,Prade1991,Dionne2006,Feigenbaum2007a,Zakharian2007,Sturman2007} for the details. The eigenvectors $(\psi_n)$ and the dispersion equation for the corresponding eigenvalues $(k_{z,n}^2)$ of \eqref{eq7} for the even \acro{TM} modes of the \acro{MIM} waveguide are
\begin{align}
	\psi_n(x)={} & H_0 \begin{cases}
										\displaystyle\frac{\cosh(\kappa_{i,n} x)}{\cosh(\kappa_{i,n} a )} & 0<x<a \\
										\me^{-\kappa_{m,n} (x-a)} & a<x<\infty
								 \end{cases}\label{field}\\
  \tanh(\kappa_{i,n} a) = {}& - \frac{\kappa_{m,n}/\epsilon_m }{\kappa_{i,n}/\epsilon_i}\label{dispersion}\\
  k_{z,n}^2 ={}& \kappa_{m,n}^2 + \omega^2 \mu \epsilon_m = \kappa_{i,n}^2 + \omega^2 \mu \epsilon_i \label{kvectors}
\end{align}
where $\Real(\kappa_{m,n}) > 0$ so that $\psi_n(x)$ does not diverge and is integrable. Here $n$ is a discrete index for the eigenvalues and the eigenfunctions. Note that we have chosen to write the modal shape in terms of the surface mode formulation of Ref. \onlinecite{Prade1991}, Sec. \acro{III}. Surface modes are the main propagating modes for the \acro{MIM} waveguide and have hyperbolic modal shapes. It is equivalently possible to describe the modes in terms of oscillatory shapes using trigonometric functions\cite{Prade1991}---analogous to the modes of the dielectric waveguide. Analytical continuation of the modal parameters $(\kappa_i, \kappa_m, k_z)$ makes the two formulations equivalent.

\begin{figure*}[htb]
\includegraphics[keepaspectratio=true,width=\textwidth]{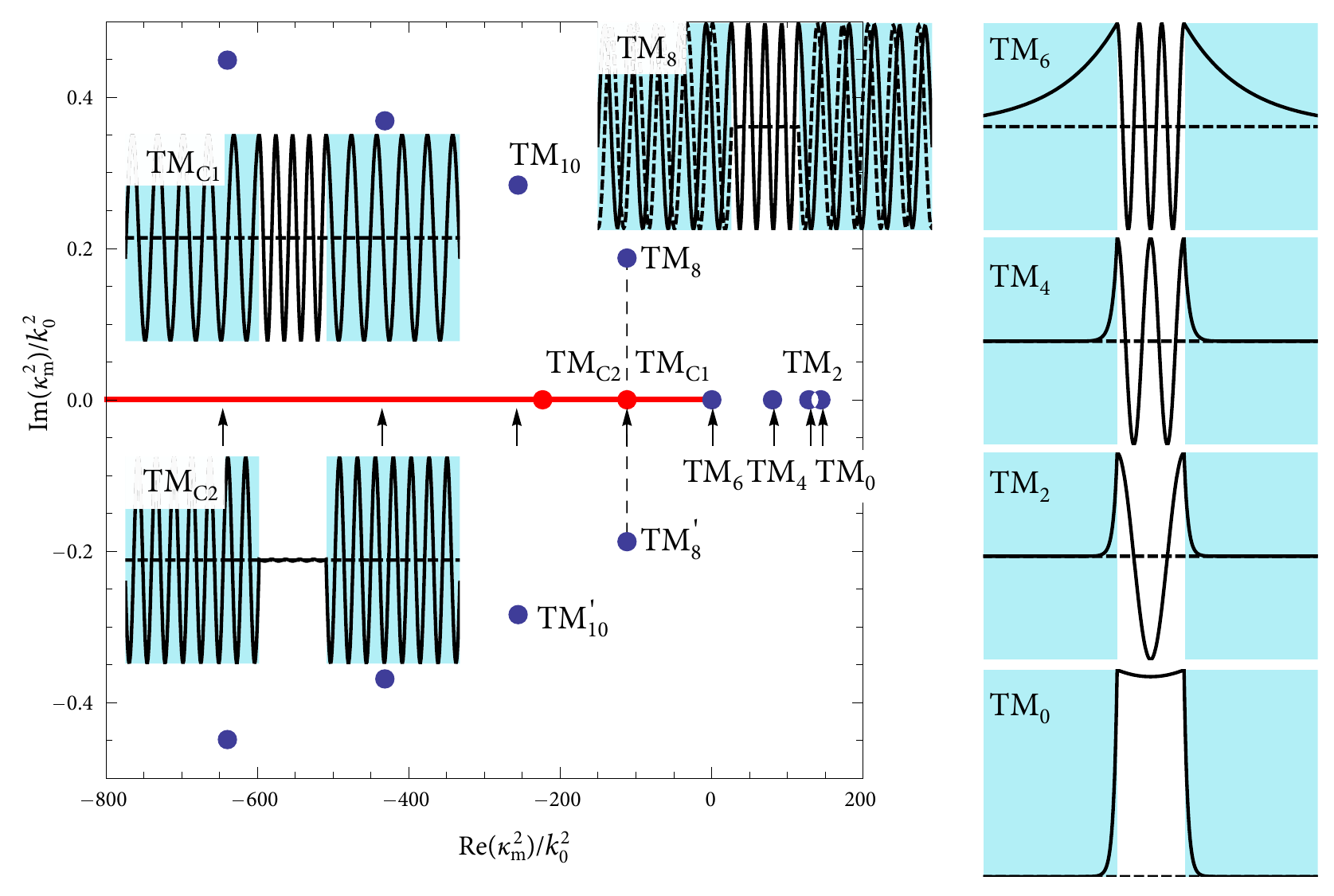}
\caption{(Color online) Spectrum of the \acro{MIM} waveguide for $\epsilon_m = -143.497$ and $2a=\lambda/4$ where $\lambda=1550\text{nm}$ is the wavelength of operation. There are four real modes and an infinite number of complex modes, all denoted with the  \textcolor{blue}{\large$\mathbf{\bullet}$} symbol. The \textbf{\textcolor{red}{thick line}} denotes the continuous spectrum. Due to the fact that $\epsilon_m$ is real, complex modes come in complex conjugate pairs. Insets show the $H_y$ mode shapes in the $x$ direction for the discrete spectrum ($\text{\acro{TM}}_0$ through $\text{\acro{TM}}_8$) and the continuous spectrum ($\text{\acro{TM}}_{\text{C1}}$ and $\text{\acro{TM}}_{\text{C2}}$)---solid lines in the insets are the real part of the mode, dashed lines are the imaginary part. The locations of the drawn continuous modes are shown by the \textcolor{red}{\large$\mathbf{\bullet}$} symbol. Modes in the continuous spectrum are purely oscillatory in the $x$ direction. Complex modes have a small decay, which is not visually apparent in the inset for $\text{\acro{TM}}_8$. Arrows $(\uparrow)$ denote the position of the modes of a parallel plate waveguide with a separation of $a$---equivalent to the limiting case $h\rightarrow0$ in Fig. \ref{fig7}.}\label{figSpectrum2D}
\end{figure*}

\begin{figure*}[ht]
\centering
\includegraphics[keepaspectratio=true,width=\textwidth]{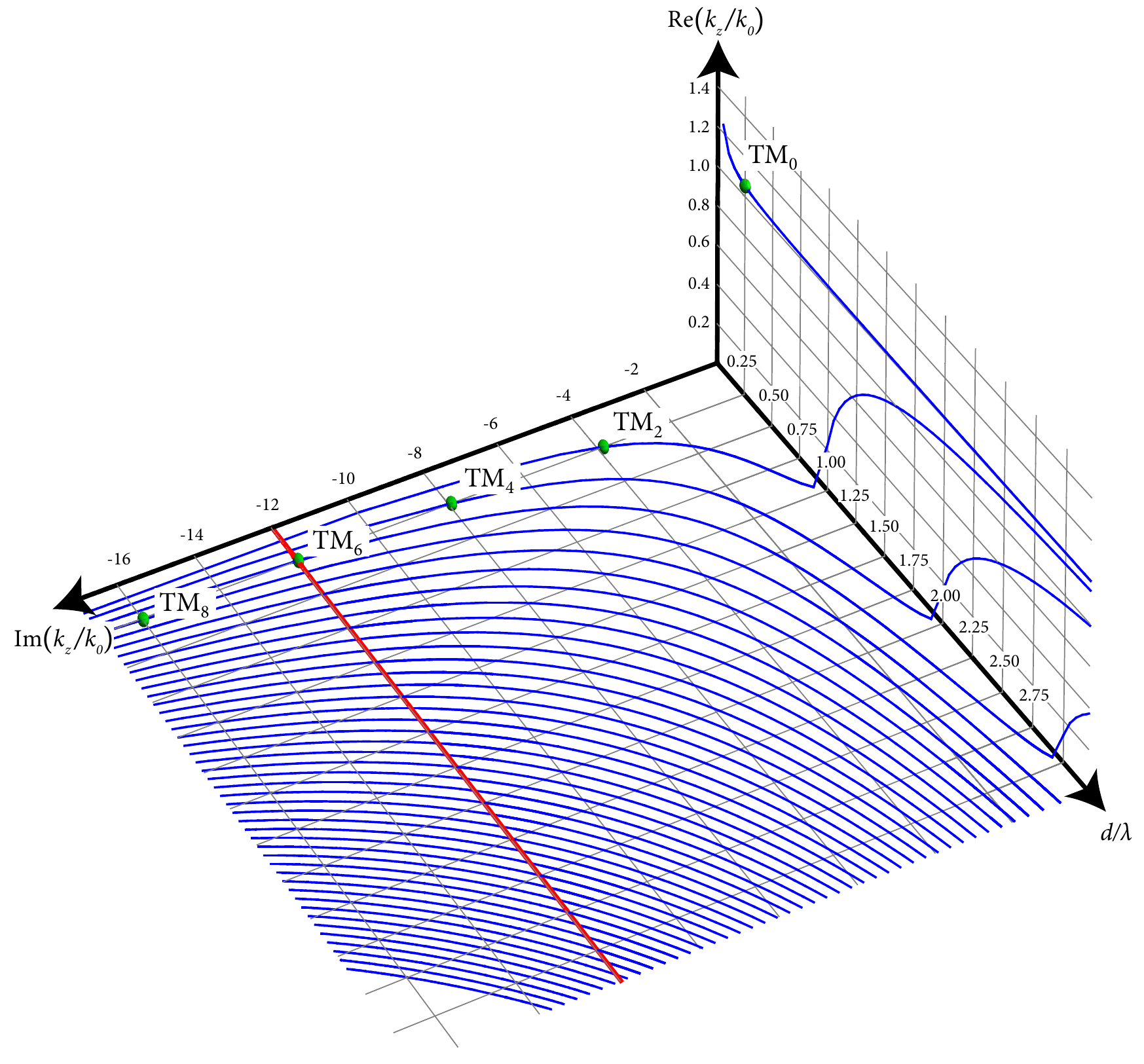}
\caption{(Color online) Spectrum of the \acro{MIM} waveguide as a function of $d=2a$, $\Real(k_z/k_0)$ and $\Imag(k_z/k_0)$ where $k_0=2\pi/\lambda$ is the free space wave vector. $\epsilon_m = -143.497$ is assumed, which corresponds to the real part of the permittivity of silver at a wavelength of $\lambda=1550\text{nm.}$ \textbf{\textcolor{red}{The thick line}} denotes the limit of the real spectrum. \textbf{\textcolor{green}{Spheres}} denote the point spectrum for the $d=2a=\lambda/4$ case as also shown in Fig. \ref{figSpectrum2D}. It can be seen that only the lowest order mode, $\text{\acro{TM}}_0$, is propagating for $d=\lambda/4$ and the rest of the modes are highly evanescent.}\label{figSpectrum3D}
\end{figure*}

\subsection{Continuous Spectrum}
In this section, we will mathematically show how a continuous spectrum can exist\cite{Zakharian2007,Sturman2007} in the \acro{MIM} waveguide and relate it to the continuous spectrum of the dielectric slab waveguide. The utility of the continuous spectrum will be evident in the mode matching analysis.

As clearly argued in Ref. \onlinecite{Shevchenko1971}, p. 16, the condition of square integrability of the modes can be replaced by the weaker condition of finiteness of the modes in their domain of definition. For the \acro{MIM} waveguide, this would imply a non-zero, yet finite electromagnetic field at infinity. These infinite-extent and, therefore, infinite energy, continuum modes (which can be normalized through the use of the Dirac delta distributions as illustrated in Ref. \onlinecite{Rozzi1997}, pp. 134--135 and Ref. \onlinecite{Miller2008}, pp. 141--148) are integrated to realize any physically possible finite energy field configuration. In this respect, such an approach is similar to the well-known Fourier transform methods, where finite energy functions are expanded in terms of the infinite energy exponentials.

Constraining fields to be finite, instead of zero, at infinity leads to the following field profile
\begin{align}
	\phi_\nu(x)={} & H_0 \begin{cases}
										\displaystyle\frac{\cosh(\kappa_{i,\nu} x)}{\cosh(\kappa_{i,\nu} a )} & 0<x<a \\
										\!\begin{aligned}
										&\cosh\left(\kappa_{m,\nu} (x-a)\right) \\
										&+ \zeta \sinh\left(\kappa_{m,\nu} (x-a)\right) 
										\end{aligned}  & a<x<\infty
								 \end{cases}\label{eq1Ba}\\
	\zeta ={}& 	\frac{\kappa_{i,\nu}/\epsilon_i}{\kappa_{m,\nu}/\epsilon_m } 
													\tanh(\kappa_{i,\nu} a) \label{eq1Bb}\\								 
  k_{z,\nu}^2 ={}& \kappa_{m,\nu}^2 + \omega^2 \mu \epsilon_m = \kappa_{i,\nu}^2 + \omega^2 \mu \epsilon_i  \label{eq1Bc}
\end{align}
which is calculated very similarly to the dielectric slab example in Ref. \onlinecite{Hanson2002}, pp. 462--470. Here $\nu$ is a continuous index for different functions in the continuous spectrum. For finite $\phi_\nu$, the arguments inside the hyperbolic functions for $x>a$ in \eqref{eq1Ba}, $\kappa_{m,\nu}$, should be purely imaginary which implies that $\Real(\kappa_{m,\nu}^2)<0$ and $\Imag(\kappa_{m,\nu}^2)=0$. These conditions can be written in terms of $k_{z,\nu}$ by using \eqref{eq1Bc} as
\begin{align*}
	\Real(k_{z,\nu}^2 - \omega^2 \mu \epsilon_m) < 0 \quad \text{and} \quad \Imag(k_{z,\nu}^2 - \omega^2 \mu \epsilon_m) = 0.
\end{align*}
Note that when \eqref{dispersion} holds true, we have $\zeta=-1$ in \eqref{eq1Bb} which makes \eqref{eq1Ba} and \eqref{field} equivalent.

\subsection{Residual Spectrum}
We saw that \eqref{eq7} is a second order differential equation which could also be written as $\op{L}H_y = k_z^2 H_y$ where $\op{L}$ is a differential operator. In Ref. \onlinecite{Friedman1990}, p. 200 it is claimed that differential operators have an empty residual spectrum, but a proof is not provided. In Ref. \onlinecite{Hanson2002}, p. 224 the residual spectrum is said not to occur in most of the electromagnetic applications, and in Ref. \onlinecite{Hanson2002}, p. 238 the residual spectrum is said to be empty for typical differential operators, though it is highlighted that such a fact is not a general result. For \eqref{eq7} we did not find any vectors belonging to the residual spectrum of $\op{L}$.

\subsection{Orthogonality relationships}
Orthogonality and completeness are two very valuable properties of modes, which make the mode matching technique, to be discussed in the next section, possible. Orthogonality of the modes for the self-adjoint, and non-self-adjoint cases are usually expressed using different definitions of the inner product. In this work, we will use the \textsl{pseudo-inner product}, $[\cdot|\cdot]$, defined as (Ref. \onlinecite{Mrozowski1997}, p. 42)
\begin{align*}
	\pseudo{f}{g} = \int_0^\infty f(x) g(x)  \dif x.
\end{align*}
It can be shown that two different eigenfunctions of $\op{L}$, $\psi_1(x)$ and $\psi_2(x)$, corresponding to two different eigenvalues $k^2_{z,1}$ and $k^2_{z,2}$ are pseudo-orthogonal with $\epsilon^{-1}(x)$ weight (Ref. \onlinecite{Chew1990},p. 330 and Ref. \onlinecite{Mrozowski1997}, p. 47)
\begin{align}
	\pseudo{\epsilon^{-1} \psi_1}{\psi_2} = 0. \label{eq17}
\end{align}
From \eqref{eq6} it can be seen that $\epsilon^{-1} \psi_1$ is proportional to the transverse electric field component $E_x$ of the mode. Therefore, the orthogonality condition can also be written as
\begin{align*}
	\int_0^\infty E_{x1}(x) H_{y2}(x)  \dif x = 
	\int_\vec{A} \vec{E}_{1}(\vec{r}) \times \vec{H}_{2}(\vec{r}) \cdot \dif \vec{A} = 0
\end{align*}
which is the well known modal orthogonality condition proved by the Lorentz reciprocity theorem (Ref. \onlinecite{Collin1991}, p. 336), where $\vec{A}$ denotes the cross section of the waveguide.

One can directly verify \eqref{eq17} by integration and using $\kappa_{m,1}^2 - \kappa_{m,2}^2 = \kappa_{i,1}^2 - \kappa_{i,2}^2$ which is a result of \eqref{kvectors}. The following orthogonality conditions between the elements of the point ($\psi_n$) and the continuous ($\phi_\nu$)  spectrum can similarly be proved
\begin{align*}
	\pseudo{\epsilon^{-1} \psi_n}{\phi_\nu} =& 0 \qquad \text{for all $n$ and $\nu$,}\\
	\pseudo{\epsilon^{-1} \phi_\mu}{\phi_\nu} =& 0 \qquad \text{for}\quad \nu\neq\mu.
\end{align*}

The orthogonality conditions talked about in this section can also be described in terms of the bi-orthogonal relationships between the eigenfunctions of the operators $\op{L}$ and $\op{L}^\dagger$ as has been done in  Ref. \onlinecite{Hanson2002}, Sec. 5.3 and Ref. \onlinecite{Kostenbauder1997}. In Ref. \onlinecite{MacCluer1991} four examples which illustrate how to choose the weight of the inner product definition so as to have orthogonal basis functions are given.

In the following sections, we will be working with fields at the junction of two different waveguides. For notational abbreviation we will use the following convention
\begin{align*}
	\field{e}{\{L,R\}}{i} = E_{x,i}^{\{L,R\}} \\
	\field{h}{\{L,R\}}{i} = H_{y,i}^{\{L,R\}}
\end{align*}
where $\{L,R\}$ is used to denote the modes of the left and right side of the junction, which leads to the following orthogonality condition
\begin{align}
	\pseudo{\field{e}{\{L,R\}}{i}}{\field{h}{\{L,R\}}{j}}=\delta_{i j} \Omega_{\{L,R\}} \label{ortho}
\end{align}
where $\delta_{i j}$ is the Kronecker delta function and $\Omega$ is the overlap integral of the electric and magnetic transverse fields.

After the classification and analysis of the \acro{MIM} waveguide modes, we will now visualize different parts of its spectrum by finding the zeros of the respective dispersion equations through the use of the argument principle method as explained in Appendix A. We will use the adjectives in Table \ref{table1} to further differentiate between the modes.
\begin{table}[ht]
	\centering
	\caption{Adjectives}
		\begin{tabular}{lcr}
		\textsl{Signifier} & &\textsl{Signified} \\ [0.5ex]
		Leaky  &  & $\Real(\kappa_m)<0$ \\
	  Proper   &  & $\Real(\kappa_m)>0$ \\
	  Improper  & & $\Real(\kappa_m)=0$ \\
	  Forward    &  &$\Real(k_z)>0$ \\
	  Backward    &  &$\Real(k_z)<0$ \\ 
		\end{tabular} \label{table1}		
\end{table}

\textsl{Leaky} modes are not normalizable and are not part of the spectrum. \textsl{Proper} modes can be normalized by the usual integration and they form the point spectrum. \textsl{Improper} modes can be normalized by using the Dirac delta functions, $\delta(x)$. They form the continuous spectrum. \textsl{Forward} modes have a positive phase velocity, whereas the \textsl{backward} modes have a negative phase velocity. We decide on the sign of $\Real(k_z)$ based on $\Imag(k_z)$: By definition, all modes are propagating in the $+z$ direction. Therefore, in the limit $z\rightarrow\infty$, the fields should go to zero. Such a behavior is possible only if $\Imag(k_z)$ is negative, since the fields have an $\exp(-\mi k_z z)$ dependence. The argument principle method gives us the $\kappa_m$ value for the modes. By using \eqref{kvectors} we get the $k_z^2$ value. We then calculate $(k_z^2)^{1/2}$ and choose the root which satisfies $\Imag(k_z)<0$. Different definitions of forward and backward modes---including the ones we use---are analyzed in Ref. \onlinecite{Shevchenko2007}. The definition we use for the leaky modes is the same as the one used in Ref. \onlinecite{Marcuse1991}, Sec. 1.5.

In Fig. \ref{figSpectrum2D} the spectrum of an  idealized lossless silver-like \acro{MIM} waveguide is shown on the plane of $\kappa_m^2$ for $\epsilon_m = -143.497$ which is the real part of the permittivity of silver at a wavelength $\lambda$ of 1550nm (Ref. \onlinecite{Hagemann1975,Palik}). There are four real modes for $2a=\lambda/4$---$\text{\acro{TM}}_0$, $\text{\acro{TM}}_2$, $\text{\acro{TM}}_4$, $\text{\acro{TM}}_6$---indexed according to the number of zero crossings in $H_y$. There is also an infinite number of complex modes, which are those with eight and more zero crossings in the insulator region. These modes have a $\kappa_m$ with a positive real part that is rather small compared to the imaginary part---this can also be deduced from the scale of the imaginary axis of Fig. \ref{figSpectrum2D}. The continuous spectrum is illustrated by the thick line which corresponds to $\Real(\kappa_m^2)<0$ and $\Imag(\kappa_m^2)=0$. This line is also the branch cut of the square root function that is used to get $\kappa_m$ from $\kappa_m^2$.

The field profiles of the modes in the insulator region, as shown in the insets of Fig. \ref{figSpectrum2D}, look quite similar to the field profile of the even modes of a parallel plate waveguide with a plate separation of $2a$. The even modes of a parallel plate waveguide have $\kappa_{i,n}^2= -n^2 \pi^2 / a^2$. We plotted the corresponding $\kappa_{m,n}^2$ values on Fig. \ref{figSpectrum2D} by using \eqref{kvectors}. It can be seen that such a description gives a quite good estimate of the location of the modes on the complex $\kappa_m^2$ plane.

$\kappa_m^2 = 0$ is the bifurcation point for the point spectrum when $\epsilon_m$ is purely real. For positive $\kappa_m^2$, the point spectrum has real modes, whereas for negative $\kappa_m^2$, the point spectrum splits into two branches that are complex conjugates of one another. $\kappa_m^2=0$ corresponds to $k_z^2=\epsilon_m k_0^2$ which then implies $k_z = -\mi \sqrt{\abs{\epsilon_m}}k_0$---bounded modes should have $\Imag(k_z)<0$. In Fig. \ref{figSpectrum3D} the point spectrum is visualized as a function of the insulator thickness $d=2a$ of the \acro{MIM} waveguide and the real and imaginary parts of $k_z$ for the same $\epsilon_m$ as in Fig. \ref{figSpectrum2D}. The thick line in Fig. \ref{figSpectrum3D} denotes the limit of the real point spectrum which is the bifurcation point of the modes. For $\Imag(k_z)<-\sqrt{\abs{\epsilon_m}}k_0$ there are no real modes. The point spectrum for the $d=2a=\lambda/4$ case of Fig. \ref{figSpectrum2D} is also highlighted with spheres in Fig. \ref{figSpectrum3D}. For the sake of clarity, we only drew one branch of modes after the bifurcation. Due to the scale of the axes in Fig. \ref{figSpectrum3D}, the quite small real part of $k_z$ for the complex modes after the bifurcation line is not visibly discernible, but is numerically there.

\begin{figure}[ht]
\centering
\includegraphics[keepaspectratio=true,width=8.6cm]{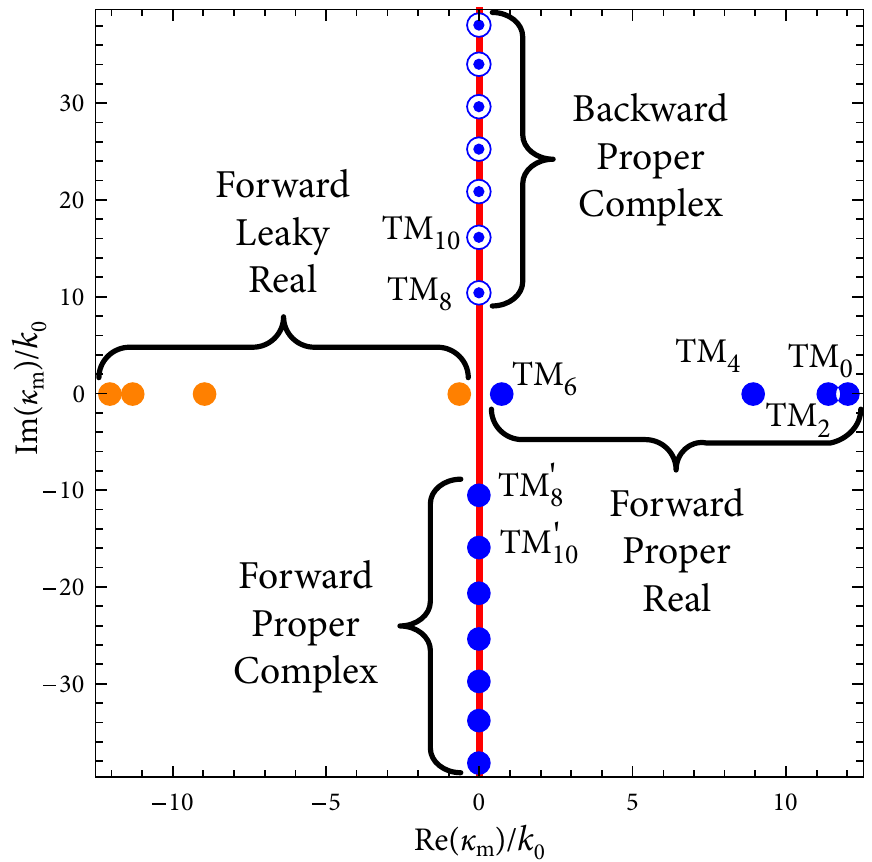}
\caption{(Color online) Visualization of the spectrum on the $\kappa_m$ plane. All parameters are the same as in Fig. \ref{figSpectrum2D}. \textbf{\textcolor{blue}{Proper}} and \textbf{\textcolor{orange}{leaky}} modes are shown for forward, {\large$\mathbf{\bullet}$}, and backward, {$\mathbf{\odot}$}, modes. The  \textbf{\textcolor{red}{thick line}} is the continuous spectrum. The complex modes have a small, yet non-zero, positive real part---which is what makes them proper modes, unlike the continuous part of the spectrum which is improper. Thus complex modes and the continuous spectrum do not intersect.}\label{fig5}
\end{figure}

\begin{figure}[ht]
\centering
\includegraphics[keepaspectratio=true,width=8.6cm]{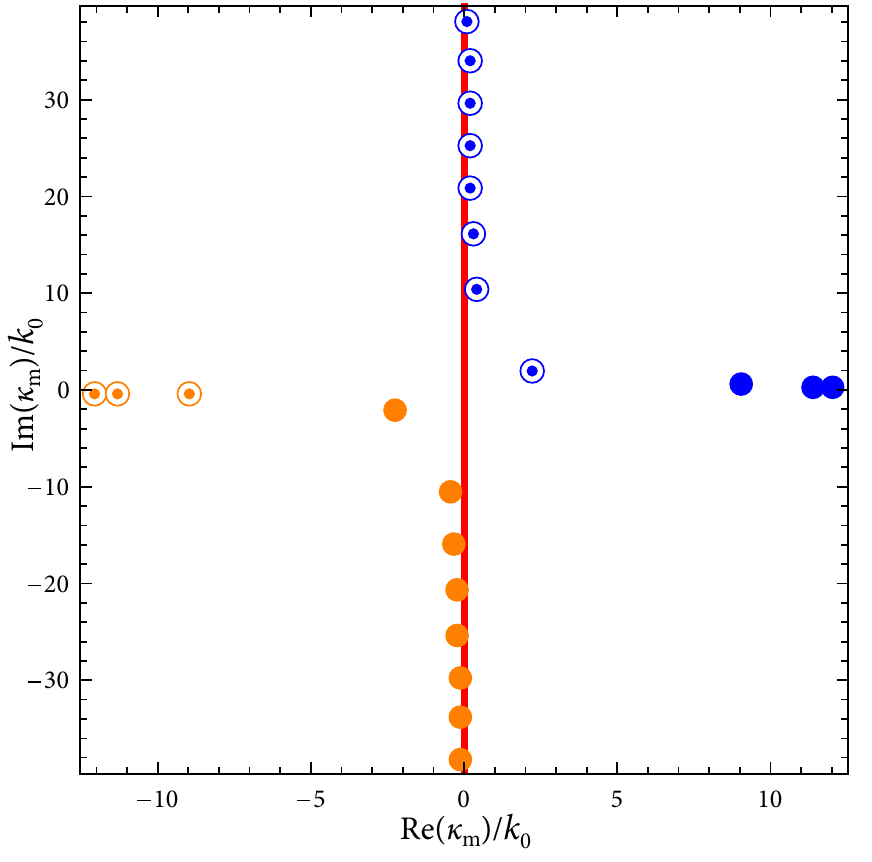}
\caption{(Color online) Effects of loss on the spectrum. All parameters are the same as in Fig. \ref{fig5} except $\epsilon_m = -143.497- \mi 9.517$. Comparison with Fig. \ref{fig5} shows that the forward, proper, complex modes on the fourth quadrant have moved to the third quadrant and thus became leaky modes. }\label{fig6}
\end{figure}

In Fig. \ref{fig5} the spectrum is plotted this time on the $\kappa_m$ plane with the same set of parameters as used in Fig. \ref{figSpectrum2D}. The two branches after the bifurcation form the forward and backward proper, complex modes. Leaky modes which are not part of the spectrum, but nevertheless are solutions to \eqref{dispersion}-\eqref{kvectors} are also shown. When loss is introduced to the metal, the spectrum moves on the complex plane as illustrated in Fig. \ref{fig6}. Forward, proper, complex modes of Fig. \ref{fig5} turn into leaky modes by migrating into the third quadrant of the complex $\kappa_m$ plane.

In Table \ref{table2} we provide the numerical values of $\kappa_m/k_0$, where $k_0=2\pi/\lambda$, for the modes in the point spectrum as labeled in Fig. \ref{figSpectrum2D}. The upper line in each row is the value for the $\epsilon_m\in\mathbb{R}$ case, the lower line is for the $\epsilon_m\in\mathbb{C}$ case. The effect of loss is greatest on the $\text{\acro{TM}}_6$ mode. 

\begin{table}[ht]
\begin{center}
\begin{minipage}{8.6cm}
\caption{$\kappa_m/k_0$ values for the modes labeled in Fig. \ref{figSpectrum2D}}
\begin{tabular}{l||D{.}{.}{21}crD{.}{.}{21}}
										& 12.02521374394057&&&\\
\rb{$\text{TM}_0$}	& 12.03170325421919&+&$\mi$& 0.39535643587620\\ \hline
										& 11.34454132059978&&\\
\rb{$\text{TM}_2$}	& 11.35226216175467&+&$\mi$& 0.41892103144838\\ \hline
										& 8.98087712606770&&\\
\rb{$\text{TM}_4$} 	& 8.99644754875734&+&$\mi$& 0.52888321538511\\ \hline
										& 0.7136870643968289&&\\
\rb{$\text{TM}_6$} 	& 2.247924588647662&+&$\mi$& 2.124681976891650\\ \hline
										& 0.00887301858491&+&$\mi$& 10.55951095636977\\
\rb{$\text{TM}_8$} 	& 0.45907739584359&+&$\mi$& 10.56851955358188\\ 
\end{tabular} \label{table2}
\end{minipage}
\end{center}
\end{table}
\section{Mode-Matching}
In this section, we will make use of the spectrum of the \acro{MIM} waveguide to calculate the scattering at the junction of two guides with different cross sections. We will use the mode matching technique\cite{Clarricoats1967} commonly used in the microwave and the optical domains\cite{Bienstman2001,Breukelaar2006,Breukelaar2006a,Oulton2007}.

\subsection{Birth of the Discretuum}
The presence of a continuous spectrum leads to the formation of integral equations when the mode-matching method is applied (Ref. \onlinecite{Rozzi1997}, Ch. 5). The integral equation is then expanded using an orthogonal basis set---not necessarily that of the modes---to solve the scattering problem.

Another way to approach the scattering problem is to limit the transverse coordinates by a \acro{PEC} wall. This approach has the effect of discretizing the continuum part of the spectrum (Ref. \onlinecite{Shevchenko1971}, pp. 38--41 and Ref. \onlinecite{Zettl2005}, pp. 204--205)---turning it into a \textsl{discretuum}\cite{footnote_4}\nocite{Bousso2000}. To limit parasitic reflections from the \acro{PEC} walls, absorbing layers can be positioned before the \acro{PEC} termination (Ref. \onlinecite{Bienstman2001}, Ch. 3). In Ref. \onlinecite{Felsen1994}, Sec. 3.2b detailed analysis of how the continuous spectrum appears from a discrete collection can be found. We will use a \acro{PEC} wall to discretize the continuous spectrum. Also, we will not use any perfectly matched layers to limit parasitic reflections since the metallic sections with permittivity $\epsilon_m$ effectively absorb the fields away from the junction.

\begin{figure}[b]
\centering
\includegraphics[keepaspectratio=true,width=8.6cm]{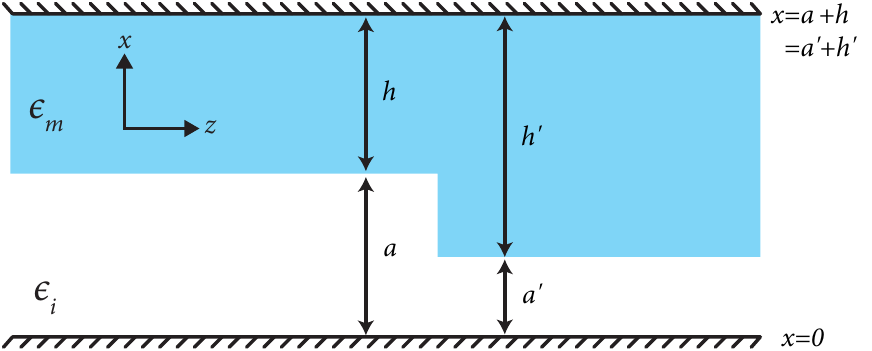}
\caption{(Color online) Geometry for mode matching. $x=a+h$ plane of Fig. \ref{fig7} is terminated by a perfect electric conductor which leads to a discretization of the continuous spectrum. } \label{fig7}
\end{figure}

The geometry is as shown in Fig. \ref{fig7}. For the left waveguide the dispersion equation for modes becomes
\begin{align}
\tanh(\kappa_{i,n} a) = {}& - \frac{\kappa_{m,n}/\epsilon_m }{\kappa_{i,n}/\epsilon_i} \tanh(\kappa_{m,n} h)\label{dispersionPEC}
\end{align}
which asymptotes to \eqref{dispersion} as $h\rightarrow\infty$. The transverse magnetic field shape is
\begin{align}
		\psi_n(x)={} & H_0 \begin{cases}
										\displaystyle\frac{\cosh(\kappa_{i,n} x)}{\cosh(\kappa_{i,n} a )} & 0<x<a \\
										\displaystyle\frac{\cosh\left(\kappa_{m,n} (x-a-h)\right)}{\cosh(\kappa_{m,n} h )} & a<x<a+h
								 \end{cases}\label{fieldPEC}
\end{align}

\begin{figure}[ht]
\centering
\includegraphics[keepaspectratio=true,width=8.6cm]{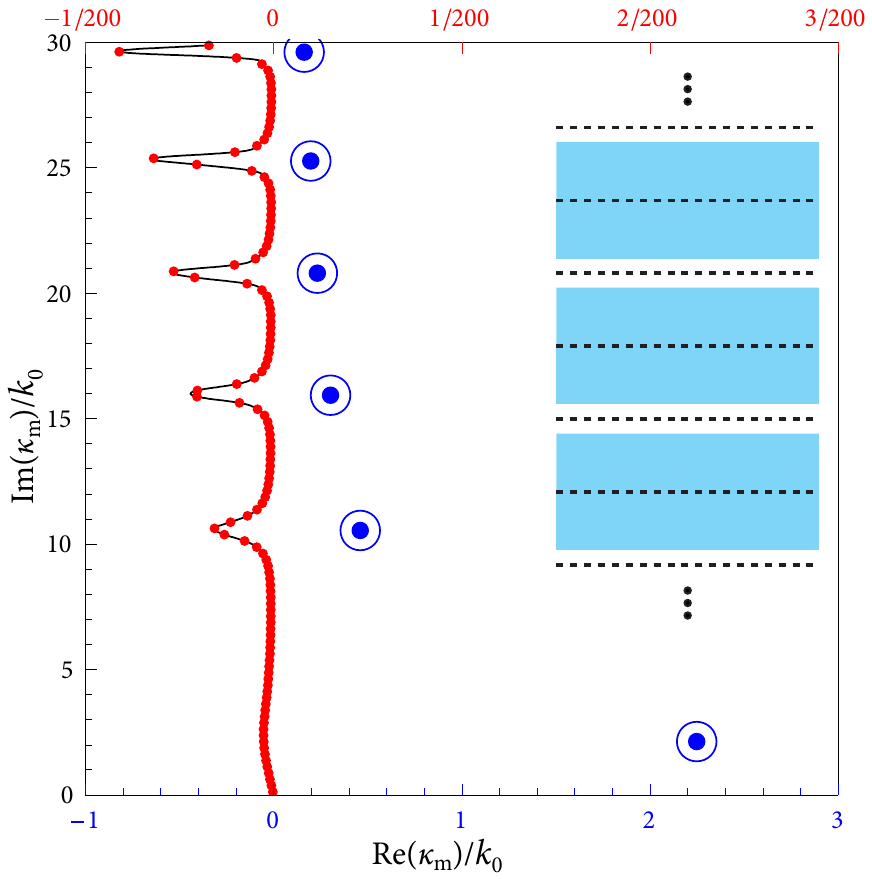}
\caption{(Color online) Effects of termination of the metal by a \acro{PEC} wall---as in Fig. \ref{fig7}---on the spectrum. $\epsilon_m = -143.497- \mi 9.517$, $\epsilon_i = 1.0$, $2a=\lambda/4$, $h=2\lambda$, $k_0=2\pi/\lambda$, $\lambda = 1550\text{nm.}$ The spectrum looks very similar to Fig. \ref{fig6}. In order to highlight the discretization of the continuum, we show a zoomed portion of the $\kappa_m$ plane. The complex modes and the discretuum show anti-crossing behavior. We use two different axes for the real part of $\kappa_m$. The scale on the bottom refers to the complex modes, \textcolor{blue}{$\mathbf{\odot}$}, whereas the one on the top refers to the discretuum, \textcolor{red}{$\mathbf{\bullet}$}. The inset shows a portion of the infinite lattice created by the repeated application of image reflection off of the \acro{PEC} boundaries. Dashed lines in the inset signify the locations of zero tangential electric field, where a \acro{PEC} termination can be applied. In separate calculations not shown in this figure, we observe that as $h$ is increased, the magnitude of the anti-crossing behavior of the continuum decreases as a result of decreased coupling between adjacent \acro{MIM} waveguides.}\label{fig8}
\end{figure}

In Fig. \ref{fig8} we show the effects of positioning a \acro{PEC} wall at the top of the lossy \acro{MIM} waveguide. The point spectrum is almost the same as in the case without a \acro{PEC} wall at the top. The continuous spectrum is discretized, and shows an anti-crossing behavior (the repulsion between modes which couple to each other) similar to the one observed in coupled waveguide systems. One way to understand the anti-crossing is to get rid of the \acro{PEC} walls by the method of images to come up with an infinite lattice of parallel \acro{MIM} waveguides each separated from each other by a distance $2(a+h)$. We observed that the perturbations to the discretuum decrease as we increase $h$, as expected from coupled mode theory. Also, the magnitude of anti-crossing behavior in the discretuum depends on the distance to the nearest mode in the point spectrum. The closer the point spectrum gets to the continuous spectrum, the larger the anti-crossing effect is. We should note that the modes of the \acro{PEC} terminated \acro{MIM} waveguide are equivalent to the even modes of a one-dimensional metallic photonic crystal at the center of the Brillouin zone as described in Ref. \onlinecite{Sturman2007}, Sec. \acro{IV}. More information on the evolution of the modes and their dependence on material properties can be found there.

\subsection{Are the modes complete?}
Before we attempt the calculation of the scattering properties of modes at waveguide junctions, we will first investigate the completeness properties of the set of modes we have at our disposal---the point and the continuous parts of the spectrum. The way we will test completeness is to expand the main mode of an \acro{MIM} waveguide of a given thickness in terms of the modes of the \acro{MIM} waveguide with a different thickness.

Suppose that we are working with the geometry depicted in Fig. \ref{fig7}. Let us expand the $k^\text{th}$ mode on the right hand side of the junction in terms of the modes on the left as
\begin{align*}
	\field{e}{R}{k}(x) = \sum_{m=1}^L A_{k m} \field{e}{L}{m}(x).
\end{align*}
In order to find $A_{k m}$ we \textsl{test} the above equation (i.e. discretize the equation by the use of integration of both sides by a given function) with $\field{h}{L}{n}$ and use \eqref{ortho} to get
\begin{align*}
	A_{k n} = \frac{\left[\field{e}{R}{k}\Big|\field{h}{L}{n}\right]}{\Omega_L^{(n)}}.
\end{align*}
Very similarly, we get the following for the magnetic fields
\begin{align}
	\field{h}{R}{k}(x) =& \sum_{m=1}^L \frac{\left[\field{e}{L}{m}\Big|\field{h}{R}{k}\right]}{\Omega_L^{(m)}} \field{h}{L}{m}(x). \label{hExpand}
\end{align}
What is the error in this expansion? We can get a measure of it by writing it as
\begin{align*}
	\field{e}{R}{k}(x) - \sum_{m=1}^L \frac{\left[\field{e}{R}{k}\Big|\field{h}{L}{m}\right]}{\Omega_L^{(m)}} \field{e}{L}{m}(x).
\end{align*}
Calculating the pseudo-inner product of the above expression with $\field{h}{R}{k}$ and then dividing it by $\Omega_R^{(k)}$ gives an error estimate as 
\begin{align*}
	\left|1 - \sum_{m=1}^L \frac{\left[\field{e}{R}{k}\Big|\field{h}{L}{m}\right] \left[\field{e}{L}{m}\Big|\field{h}{R}{k}\right]}{\Omega_L^{(m)} \Omega_R^{(k)} }\right|.
\end{align*}
Calculating the error based on the magnetic field expansion results in the same expression. 

\begin{figure}[ht]
\centering
\includegraphics[keepaspectratio=true,width=8.6cm]{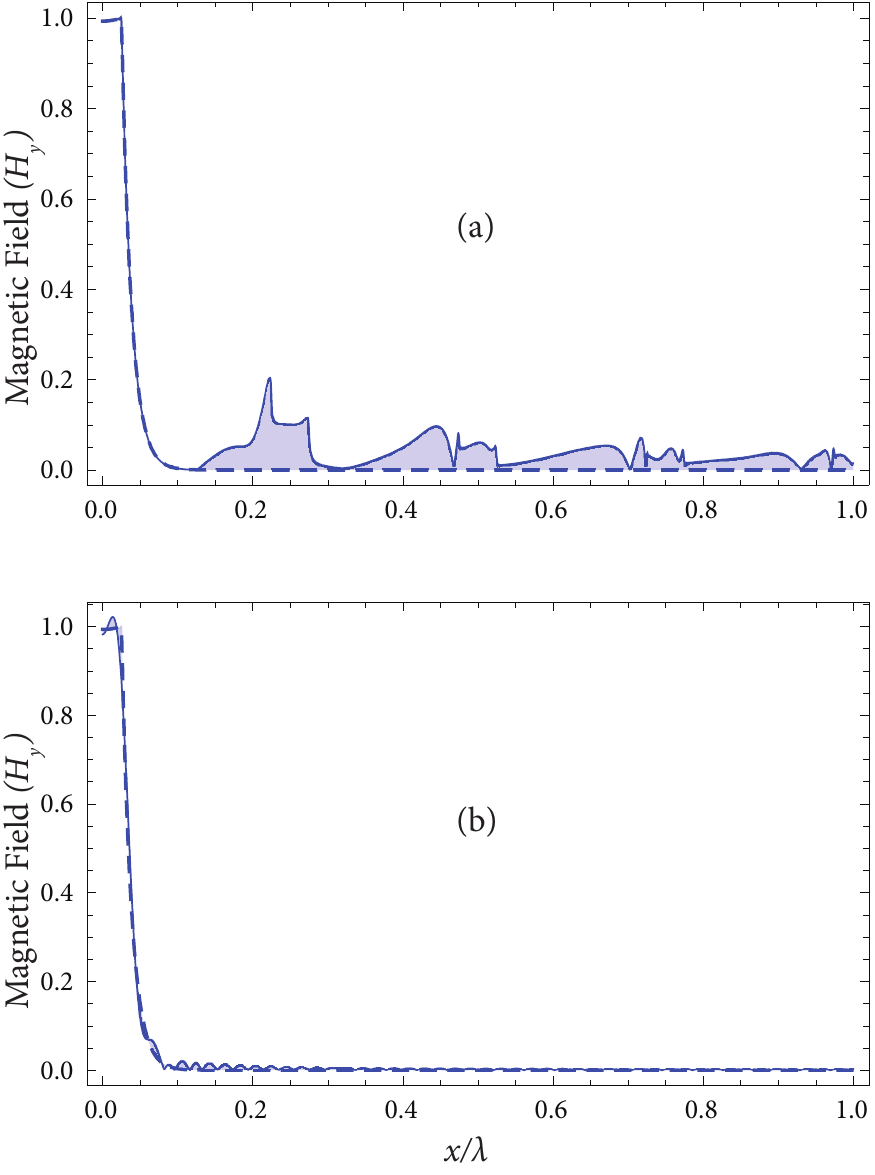}
\caption{(Color online) Effects of the continuous spectrum on mode expansion. $\epsilon_m = -143.497- \mi 9.517$. The main mode of the right waveguide, $\field{h}{R}{1}(x)$, with $2a^\prime=\lambda/20$ is expanded in terms of the modes of the left waveguide, $\field{h}{L}{m}(x)$, with $2a=\lambda/4$. The dashed lines show the magnitude of the magnetic field profile of the main mode of the right waveguide. The solid line shows the result of the expansion of \eqref{hExpand}. Shaded regions show the error from the use of a particular expansion, i.e., the difference between the expansion and the actual right waveguide mode. Due to symmetry, only half of the full modal profile is plotted. (a) Open waveguide of Fig. \ref{figGeometry}, expansion made using the point spectrum only. (b) \acro{PEC} terminated \acro{MIM} waveguide of Fig. \ref{fig7}, with $a+h=a^\prime+h^\prime=\lambda$, expansion made using the discretuum and the point spectrum.}\label{fig10}
\end{figure}

In Refs. \onlinecite{Zaki1988,Omar1987,Omar1985,Omar1986,Sturman2007a,Sturman2008} the importance of the complex modes has been demonstrated. In Fig. \ref{fig10} we show the importance of the continuous spectrum. As shown in Fig. \ref{fig10}(a) without the continuous spectrum, the field expansion converges, but to a field profile which is not the same as the desired profile of the right junction. On the other hand, inclusion of the continuous spectrum through the discretization of the continuum by a \acro{PEC} wall leads to the correct field profile as illustrated in Fig. \ref{fig10}(b). It can be observed that the expansion based on the point spectrum only quite nicely fits the field profile of the right waveguide in the insulating region of the left waveguide $(x/\lambda<0.125)$; however, in the metal region $(x/\lambda>0.125)$ the expansion fails. The point spectrum of the left waveguide has an exponentially decaying field profile for $x/\lambda>0.125$ which turns out to be insufficient for the expansion of an arbitrary field profile in the metal. The continuous spectrum, with its non-decaying field profile, makes field expansion in the metal region possible.

\subsection{Field Stitching}
Now that we know how to treat the continuous spectrum and are confident that the collection of the point and the continuous spectrum results in a complete basis set, we can proceed with the mode-matching formalism. We will begin by assuming that the $p^\text{th}$ mode of the left waveguide propagates toward the right, scatters and creates the following set of fields at the right and left sides of the junction, which by the continuity of the tangential magnetic and electric fields, are set equal
\begin{align}
	\sum_{m=1}^{\infty} \left(\delta_{m p} + R_{m p}\right)\field{h}{L}{m}(x) ={}& 
			\sum_{k=1}^\infty T_{k p} \field{h}{R}{k}(x) \label{junctionH}\\
	\sum_{m=1}^{\infty} \left(\delta_{mp} - R_{mp}\right)\field{e}{L}{m}(x) ={}& \sum_{k=1}^\infty T_{k p} 
			\field{e}{R}{k}(x).\label{junctionE}
\end{align}
Here $R_{m p}$ is the reflection coefficient of the $m^{\text{th}}$ mode of the left waveguide in response to an incoming field in the $p^{\text{th}}$ mode. Similarly, $T_{k p}$ is the transmission coefficient of the $k^\text{th}$ mode of the right waveguide. Note that we chose $R_{m p}$ to denote the reflection coefficient for the transverse magnetic fields, which automatically results in $-R_{m p}$ as the reflection coefficient for the transverse electric fields.

In Ref. \onlinecite{Eleftheriades1994}, it is shown that the testing of the above equations should be done by the magnetic field of the larger waveguide for enforcing electric field continuity \eqref{junctionE} and by the electric field of the smaller waveguide to enforce the magnetic field continuity \eqref{junctionH}. Although that analysis was specifically done for waveguides with perfect metals $(\abs{\epsilon_m}\rightarrow\infty)$, we still use that strategy so that the formulation limits to the correct one should the metals be made perfect. 

For those cases where $a<a^\prime$, we will take the pseudo-inner product of \eqref{junctionH} with $\field{e}{L}{n}$ and of \eqref{junctionE} with  $\field{h}{R}{n}$. Furthermore, assuming there are $L$ modes on the left and $R$ modes on the right, we get
\begin{align*}
	\sum_{m=1}^L \left(\delta_{m p} + R_{m p}\right)\Omega_L^{(m)} \delta_{m n} ={}& 
			\sum_{k=1}^R T_{k p} \left[\field{e}{L}{n}\Big|\field{h}{R}{k}\right] \\
	\sum_{m=1}^L \left(\delta_{mp} - R_{mp}\right)\left[\field{e}{L}{m}\Big|\field{h}{R}{n}\right] ={}& 
			\sum_{k=1}^R T_{k p} \Omega_R^{(k)} \delta_{k n}
\end{align*}
with the help of \eqref{ortho}. When $a>a^\prime$ by using $\field{e}{R}{n}$ and $\field{h}{L}{n}$ to test \eqref{junctionH} and \eqref{junctionE} respectively, we arrive at the following set of equations
\begin{align*}
	\sum_{m=1}^L \left(\delta_{mp} + R_{mp}\right)\left[\field{e}{R}{n}\Big|\field{h}{L}{m}\right] ={}& 
			\sum_{k=1}^R T_{k p} \Omega_R^{(k)} \delta_{k n} \\			
	\sum_{m=1}^L \left(\delta_{m p} - R_{m p}\right)\Omega_L^{(m)} \delta_{m n} ={}& 
			\sum_{k=1}^R T_{k p} \left[\field{e}{R}{k}\Big|\field{h}{L}{n}\right].	
\end{align*}
These are linear matrix equations with $R_{m p}$ and $T_{k p}$ as the unknowns. After calculating the inner products, the set of equations can be inverted to give the reflection and transmission coefficients for the modes.

\begin{figure}[ht]
\centering
\includegraphics[keepaspectratio=true,width=8.6cm]{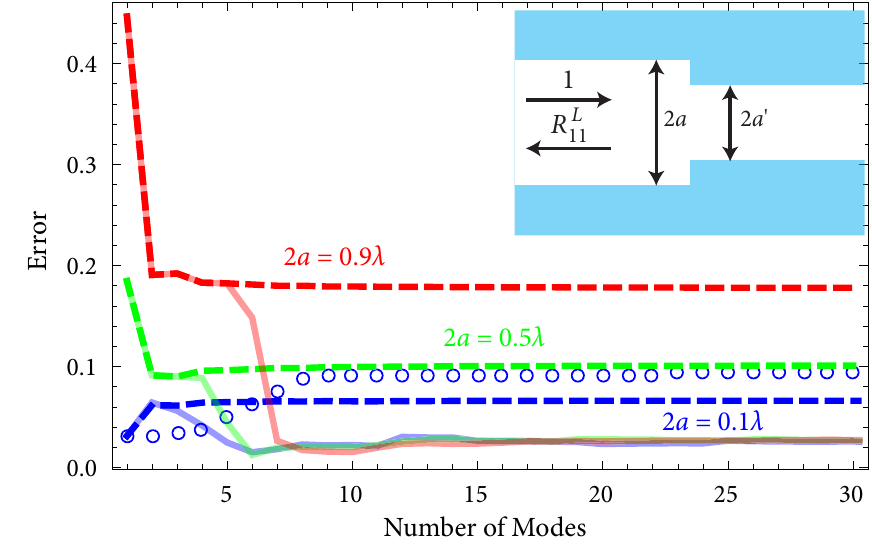}
\caption{(Color online) Convergence study of the reflection coefficient, $R_{11}^L$, of the main mode of the left waveguide traveling toward the right waveguide for \textcolor{red}{$\mathbf{2a = 0.9\lambda}$}, \textcolor{green}{$\mathbf{2a = 0.5\lambda}$} and \textcolor{blue}{$\mathbf{2a = 0.1\lambda}$}. $a^\prime/a = 0.4$ and $\epsilon_m=-143.497-\mi9.517$ for all cases. Dashed lines are for calculations including the point spectrum only. Solid lines are those with both the point and the continuous spectrum. Empty circles denote the calculations done with the \textsl{forward} point spectrum and the continuous spectrum for the \textcolor{blue}{$\mathbf{2a = 0.1\lambda}$} case only. Error is defined as $\abs{(R_{11}^{L,\text{\acro{MM}}}-R_{11}^{L,\text{\acro{FDFD}}})/R_{11}^{L,\text{\acro{FDFD}}}}$ where \acro{MM} stands for mode-matching and \acro{FDFD} for finite-difference frequency-domain calculations. The inset shows the junction geometry.}\label{fig11}
\end{figure}

In Fig. \ref{fig11}, we compare the mode-matching method with the \textsl{finite-difference frequency-domain} (\acro{FDFD}) technique \cite{Veronis2005}. In Ref. \onlinecite{Kocabas2008} scattering at \acro{MIM} junctions was investigated using \acro{FDFD}. It takes relatively few modes for the mode-matching calculations to converge. Without the continuous spectrum, the mode matching results converge to the wrong result. Inclusion of the continuous spectrum decreases the error to around 2\%, which is probably due to the space discretization of \acro{FDFD} simulations as well as the method used in the de-embedding of the scattering coefficients from fields. As is also evident from Fig. \ref{fig11} the utility of the single mode $(L=R=1)$ mode-matching calculations increases as the dimensions of the waveguides decrease. The single mode approximation is closely related to the simplified impedance model investigated in Ref. \onlinecite{Kocabas2008} where it was shown that for deep subwavelength structures impedance models are a good approximation. In Fig. \ref{fig11} we also show the effect of neglecting the backward modes in the mode-matching calculations for the $2a=0.1\lambda$ case. Backward modes are important in this sub-wavelength geometry; however, for the wider geometries of the $2a=0.5\lambda$ and $2a=0.9\lambda$ cases we did not observe any increase in the error when backward modes were neglected in the mode-matching calculations.

\begin{figure}[ht]
\centering
\includegraphics[keepaspectratio=true,width=8.6cm]{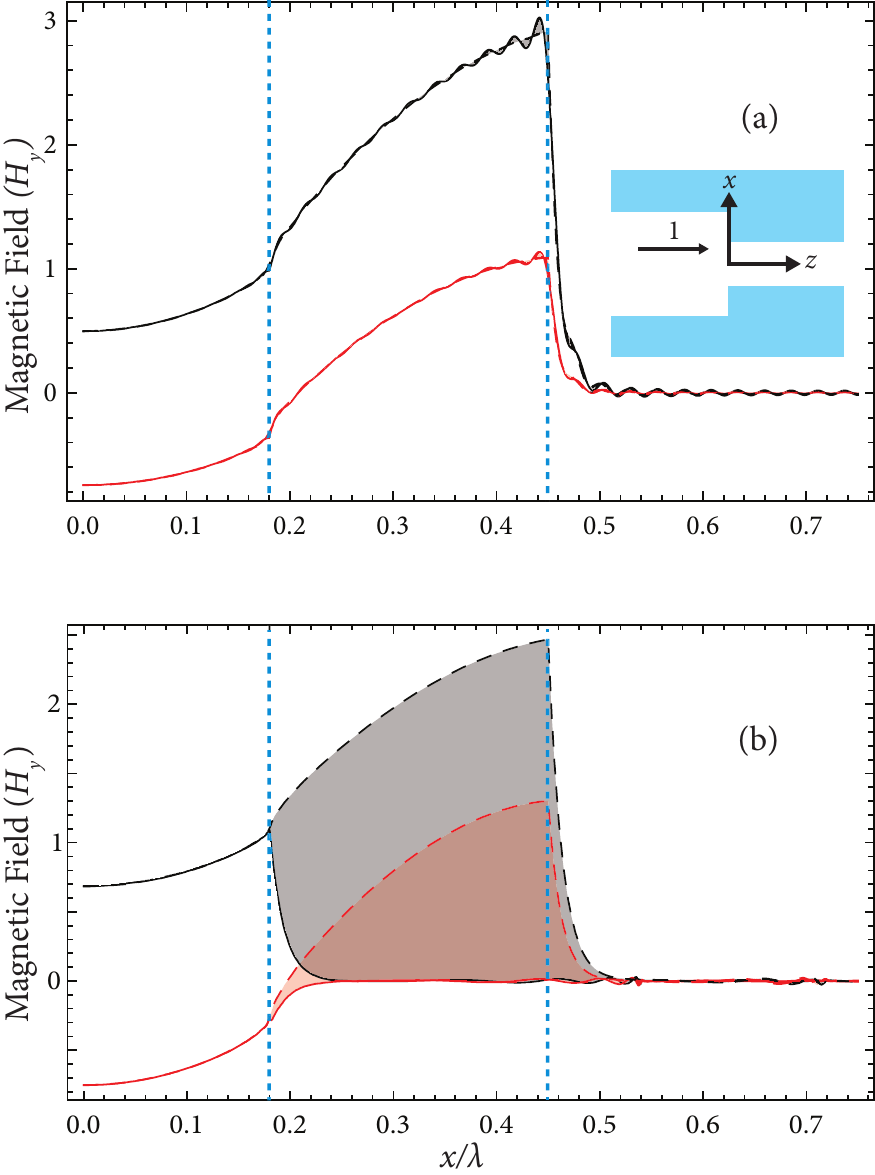}
\caption{(Color online) Magnetic field $H_y$ at the junction of two \acro{MIM} waveguides as a result of the scattering of the main mode of the left waveguide traveling toward the right---the inset shows the schematic of the geometry. The left waveguide has $2a=0.9\lambda$, the right waveguide has $2a^\prime=0.36\lambda$ so that $a^\prime/a=0.4$ as in Fig. \ref{fig11}. $a+h=a^\prime+h^\prime=3\lambda/4$. Fields on the left of the junction are dashed, fields on the right are shown with solid lines for both the \textbf{real} and the \textbf{\textcolor{red}{imaginary}} part of the magnetic field profile. The difference between the left and right fields is shaded. Due to symmetry, only half of the field profile is plotted. Vertical dotted lines at $x=0.18\lambda$ and $x=0.45\lambda$ denote the end of the insulator region for the right and the left side of the junction respectively. (a) Mode-matching calculations using the point and the continuous spectrum---60 modes in total---showing good agreement between the fields just on the left and just on the right of the junction. (b) Mode-matching calculations using the point spectrum only---100 modes in total---showing clear disagreement between the calculated fields just on the left and just on the right of the junction.}\label{fig12}
\end{figure}

Analysis of the convergence of the field expansions on both sides of a junction is an important criterion for assessing the validity of the mode-matching technique\cite{Bhattacharyya2003,Lee1971,Leroy1983,Mittra1972}. In Fig. \ref{fig12} we show the magnetic field profile at the junction of two \acro{MIM} waveguides. As is evident from the figure, convergence of the fields on both sides of the junction is obtained only when the continuous spectrum is also taken into consideration.  Otherwise, the fields just on the left and just on the right of the junction do not agree with one another, showing one or both calculations to be in error. The clear conclusion from this numerical illustration is that the point spectrum on its own is not sufficient to describe the behavior of the waveguide junctions. Inclusion of the continuous spectrum is essential.

\begin{figure}[ht]
\includegraphics[keepaspectratio=true,width=8.6cm]{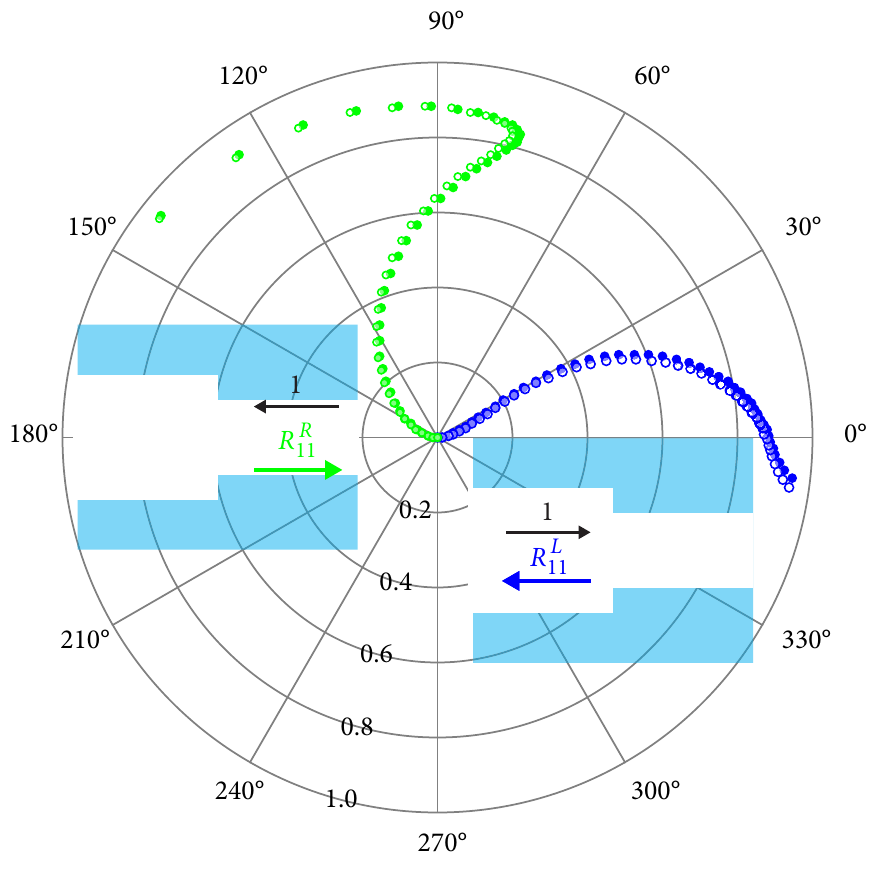}
\caption{(Color online) Reflection coefficient of the main mode at the junction between two \acro{MIM} waveguides of different insulator thicknesses, plotted on the complex plane within the unit circle. Both waveguides have $\epsilon_m = -143.497- \mi 9.517$ and $\epsilon_i=1.0$. Filled circles, {\large$\mathbf{\bullet}$}, are \acro{FDFD} results, empty circles, {\large$\mathbf{\circ}$}, are mode-matching results. The left waveguide thickness is fixed at $2a=0.9\lambda$. The right waveguide thickness varies from $2a^\prime=\{0.02\lambda, 0.04\lambda,\dots,0.9\lambda\}$. The origin is the zero reflection point that corresponds to $a=a'$. As $a'$ decreases progressively toward zero, we move progressively along the curves away from the origin. The \textbf{\textcolor{blue}{first set of curves}}, $R_{11}^L$, are for the  case when the \textsl{mode of the left waveguide}, traveling from left to right, is scattered by the junction. The \textbf{\textcolor{green}{second set of curves}}, $R_{11}^R$, are for the case when the main \textsl{mode of the right waveguide}, traveling from right to left, is scattered by the junction. Insets illustrate the respective cases. Details about the \acro{FDFD} calculations can be found in Ref. \onlinecite{Kocabas2008}.}\label{figFDFDcompare}
\end{figure}

In Fig. \ref{figFDFDcompare} we visualize the scattering coefficient of the main mode of the \acro{MIM} waveguide. We do the calculations in two different ways, one using \acro{FDFD}, and the other using the mode-matching technique with the point and the continuous spectrum. When applying mode-matching, we use the $a>a^\prime$ formulation for $R_{11}^L$ calculations and the $a<a^\prime$ one for $R_{11}^R$. There is a very good match between the results of the two techniques, verifying the applicability of the mode-matching method.

\section{Discussion}
We will begin this section by a comparative study of the modal structures for the parallel plate, the dielectric slab and the \acro{MIM} waveguides. Our aim will be to frame the \acro{MIM} case as a bridge between the dielectric slab and the parallel plate waveguides.

The two dimensional symmetric dielectric slab waveguide---where the cladding and the core are composed of lossless, positive permittivities---has no proper complex modes (Ref. \onlinecite{Collin1991}, p. 718) as expected from the regular Sturm-Liouville theory. The dispersion equation for the  dielectric slab waveguide---which is the same as \eqref{dispersion}---does have complex roots. However, all of those complex roots correspond to leaky modes that are not a part of the spectrum.

The point spectrum of the dielectric slab consists of real propagating modes with pure exponential decay in the cladding region and with positive power flux in the direction of propagation, $z$. The number of modes in the point spectrum is a finite quantity. The cutoff condition for the point spectrum is obtained by the equality $\kappa_m^2=0$. The continuous spectrum of the dielectric slab starts just below the cutoff condition for the point spectrum where $\kappa_m^2<0$. In this range, the modes extend to infinity in the transverse direction, but they remain bounded. We can divide the continuous spectrum into two sections based on the sign\cite{footnote_5} of $k_z^2 = \kappa_m^2 + \omega^2 \mu_0 \epsilon_m = \kappa_m^2 + \omega^2 \mu_0 |\epsilon_m|$. The section where $k_z^2 >0$ is called the  \textsl{radiative part} whereas the section $k_z^2 <0$ is called the \textsl{reactive part} of the continuous spectrum (Ref. \onlinecite{Rozzi1997},pp. 128--132 and Ref. \onlinecite{Marcuse1991}, pp. 19--28). For both the radiative and the reactive parts of the continuous spectrum, the transverse field profiles of the modes are sinusoidal standing wave patterns. As the sinusoidal variation in the transverse direction $(\kappa_m = \mi |\kappa_m|)$ becomes more rapid, the sinusoidal variation in the propagation direction ($k_z$) decreases. The jump from the radiative to the reactive part occurs when the sinusoidal variation in $z$ goes to zero: $k_z = 0$, $\kappa_m/k_0 = \mi \sqrt{|\epsilon_m|}$. In the reactive part, the variation in $x$ is so rapid that, modes decay as they propagate in $z$. The radiative part is often visualized in the mind's eye by plane waves that originate in the cladding region far away from the core, propagate at an angle towards the core, reflect off of the core and propagate away from it. The interference pattern between the incoming and the outgoing plane waves leads to a standing wave pattern in the transverse direction and a propagating plane wave in the positive $z$ direction. The reactive part is harder to think about in terms of plane wave propagation since in this part of the continuous spectrum $k_z$ is purely imaginary and, therefore, the modes are decaying in the positive $z$ direction. The plane wave picture is often extended to the reactive range by allowing for the possibility for the plane waves to come at an `imaginary' angle of incidence. 

The spectrum of the parallel-plate waveguide with \acro{PEC} boundaries is less nuanced than that of the dielectric slab case. There is no continuous part of the spectrum. All modes belong to the point spectrum and form a discrete basis set. There are infinitely many discrete modes supported by the parallel-plate waveguide. Finitely many of them are propagating modes and carry a positive energy flux in the $z$ direction. The remaining ones do not carry any energy and are called \textsl{evanescent}.

The spectrum of the \acro{MIM} waveguide is a hybrid of the dielectric slab and the parallel-plate waveguides' spectra. The \acro{MIM} waveguide has both the point and the continuous spectra. Let us start with the continuous spectrum first.

In Section \acro{III} we have shown that the continuous spectrum of the lossless \acro{MIM} waveguide exists for all $k_z^2 < \omega^2 \mu \epsilon_m$ which is equivalent to $k_z^2 < -\omega^2 \mu |\epsilon_m|$ since $\epsilon_m$ is a large negative number for the \acro{MIM} case. We see that the continuous spectrum of the \acro{MIM} waveguide is purely reactive and can be thought of as composed of plane waves coming at an imaginary angle through the metal and reflecting off of the insulator region. Equivalently, at a more fundamental level, they are the solutions of the wave equation with the condition that the fields be finite and non-zero at infinity. 

The point spectrum of the \acro{MIM} waveguide has infinitely many members, similar to the parallel-plate case. Indeed, one can think of the point spectra of the \acro{MIM} and the parallel-plate waveguides as analytical continuations of each other as illustrated in Ref. \onlinecite{Takano1972} for the main $\text{\acro{TM}}_0$ mode of the \acro{MIM} and the \acro{TEM} mode of the parallel-plate waveguides. Consequently, the main $\text{\acro{TM}}_0$ mode of the \acro{MIM} waveguide can be thought of as the symmetrical coupling of the two surface plasmon modes at the top and the bottom metal-insulator interfaces\cite{Prodan2003}. 

There are no evanescent discrete modes in the dielectric slab waveguide, but the \acro{MIM} waveguide supports them. As illustrated in Fig. \ref{figSpectrum3D} there are infinitely many discrete modes for any given insulator thickness. Finitely many of those are real modes $(k_z^2 \in \mathbb{R})$, the remaining ones are complex $(k_z^2 \in \mathbb{C})$. Of the real modes, only those with $k_z \in \mathbb{R}$ carry any power flux. These observations are strictly true only for the case $\epsilon_m \in \mathbb{R}$. When there is loss in the system, all modes do carry small, yet finite, amount of power. We have verified these claims by calculating the power flux as
\begin{widetext}
\begin{align*}
P=\frac{1}{2}\Real\left(\int E_x(x) H_y^*(x) \dif x \right) =\frac{1}{2}|H_0^2| \times \Real \Bigg\{ \frac{k_z}{\omega \epsilon_m} \frac{1}{\Real(\kappa_m)} + 
\frac{k_z}{\omega \epsilon_i}
			\frac{\sinh(\kappa_{i\scriptscriptstyle\mathbb{R}} d)\big/\kappa_{i\scriptscriptstyle\mathbb{R}}
						+\sin(\kappa_{i\scriptscriptstyle\mathbb{I}} d)\big/\kappa_{i\scriptscriptstyle\mathbb{I}}}
					 {\cosh(\kappa_{i\scriptscriptstyle\mathbb{R}} d)+\cos(\kappa_{i\scriptscriptstyle\mathbb{I}}d)} \Bigg\}
\end{align*}
\end{widetext}
using \eqref{eq6}, \eqref{field} and $\kappa_i = \kappa_{i\scriptscriptstyle\mathbb{R}} + \mi \kappa_{i\scriptscriptstyle\mathbb{I}}$ for different modes in the point spectrum. For waveguide junctions, it has been argued that neglecting complex mode pairs on either side of a waveguide junction leads to a discontinuity in the reactive energy stored at the junctions\cite{Omar1986}. A similar argument can also be made for the highly evanescent continuous modes of the \acro{MIM} geometry. Real and complex bound modes are exponentially decaying in the metal region. At the junction between two \acro{MIM} waveguides, the smaller waveguide's discrete bound modes cannot account for the field leakage into---and therefore reactive energy storage in---the metal region due their exponential decay. The continuous modes which extend infinitely into the metal region make it possible to account for the leakage into metal regions.

We have shown the necessity to take into account the full modal structure of the \acro{MIM} waveguide by the calculations we presented in Section \acro{IV}. Here, we should note that, when we did mode-matching calculations for the lossless \acro{MIM} geometry, we occasionally observed convergence problems while we were trying to reproduce the \acro{FDFD} results. However, when we included loss, all our calculations converged and we did reproduce the lossy \acro{FDFD} results as we have illustrated in Fig. \ref{figFDFDcompare}. 
The matrix that one needs to invert to solve the mode-matching equations has a higher condition number in the lossless case compared to the lossy one. That may explain the difficulties we faced.

In the remainder of this section, we will draw some connections between optics and other branches of science with the hope of expanding the analogical toolset we use for analysis---as exemplified in Ref. \onlinecite{Kay1956}. 

The one dimensional Schr{\"o}dinger equation and the electromagnetic wave equations in layered media are closely related. Both are in the Sturm-Liouville form and one can map the dispersion equation for the \acro{TE} mode of a dielectric slab waveguide to the dispersion equation for the modes of a finite potential well (Ref. \onlinecite{Lekner1987}, p. 11). Furthermore, if one allows for the finite potential well to have different effective masses in the well and the barrier regions, then a mapping to the \acro{TM} mode dispersion equation \eqref{dispersion} also becomes possible. In Refs. \onlinecite{Siewert1978,Paul2000,Blumel2005} exact closed form analytical solutions for the modes of a single, finite quantum well are developed. These solutions can be mapped to the transverse electric modes of the \acro{MIM} or the dielectric slab waveguides and perhaps with some labor could be expanded to the \acro{TM} case as well. Effects of discontinuities in quantum well potentials are shown to lead to changes in the reflection spectrum of the wells in Ref. \onlinecite{Ahmed2000}. It is intriguing to ask whether such studies could be useful in optics for the investigation of the effects of material interfaces. Recently, it was shown that non-hermitian potentials in the Schr\"odinger equation can have purely real spectra due to the certain symmetries of the Hamiltonian of the system\cite{Bender2007}. Pseudo-hermiticity, which we have touched upon in Section \acro{III}, has been shown to play an important role in the interpretation of these systems\cite{Geyer2008}. In Ref. \onlinecite{Krejcirik2008} a parallel plate waveguide with impedance boundary conditions was analyzed by the help of the definition of an inner product which reveals some hidden symmetries\cite{Albeverio2008,Kuzhel2008} of the system. The operator theoretic findings summarized in Ref. \onlinecite{Bender2007} can have implications for the analysis of waveguides.

Lastly, in the microwave literature, the unique definition---if there is any---of the impedance of an arbitrary waveguide mode is an active area of research. The causal waveguide impedance definition of Ref. \onlinecite{Williams2003} seems to formulate a unifying framework to merge different interpretations together. It seems worthwhile to ask what the $\text{\acro{TM}}_0$ mode impedance of an \acro{MIM} waveguide would be for a causal definition of the impedance given the Kramers-Kronig relationships for waveguide modes as investigated in Ref. \onlinecite{Haakestad2005}.

\section{Conclusion}
In this paper we investigated the even \acro{TM} modes that the \acro{MIM} waveguide supports. We based our  analysis in the language of operators and used methods developed for Sturm-Liouville systems to expand the results reported in Ref. \onlinecite{Sturman2007}. The mathematical structure of the odd modes are very similar to the even ones and can be derived in a similar manner. These findings were in accordance with Ref. \onlinecite{Jablonski1994} where it was shown that in general, open structures will have complex and continuous spectra. 

After the investigation of the modes, we showed their utility and relevance by the mode-matching method. We investigated the problem of modal scattering at the symmetric junction of two \acro{MIM} waveguides with different cross sections\cite{Kocabas2008} and successfully applied the mode-matching technique to predict the modal reflection coefficients calculated by full-field simulations. Lastly, we commented on some of the possible links between the quantum mechanics, optics and microwave literature and considered possible research directions.

The knowledge of the set of orthogonal modes which form a complete basis for a given geometry leads to a much more simplified algebra and speeds up calculations. The results of this paper are valuable for electromagnetic scattering calculations involving the \acro{MIM} geometry. Our results would also help in the analysis of optics experiments involving \acro{MIM} waveguides \cite{Lopez-tejeira2007,Verhagen2008a,Dionne2008}. Furthermore, the results reported are also useful for analyzing plasmonic quantum optics \cite{Klimov2004,Chang2006,Chang2007,Jun2008} and Casimir effect devices\cite{Intravaia2005,Capasso2007}.

The analysis made in this paper can be generalized for other related geometries involving metals at optical frequencies\cite{Collin2007,Kong2007,Kong2007a,Vogel2007}. The rich set of modes available in the \acro{MIM} geometry suggests that modal investigation of arbitrary three-dimensional nano-metallic waveguides---which are thought to replace the electrical interconnects on future computing devices---will require novel means of deducing their discrete and continuous spectra. 

\begin{acknowledgments}
This work is supported by a seed grant from \acro{DARPA} \acro{MTO}, the Interconnect Focus Center, one of five research centers funded under the Focus Center Research Program, a \acro{DARPA} and Semiconductor Research Corporation program, and  the \acro{AFOSR} ``Plasmon Enabled Nanophotonic Circuits'' \acro{MURI} Program.
\end{acknowledgments}

\appendix
\section{Argument Principle Method}
In this section we will briefly describe the method we used to find the zeros of \eqref{dispersion} and \eqref{dispersionPEC}. One of the main of problems of any root finding algorithm is the starting point in the search domain. When the search needs to be done over two dimensions of the complex plane, brute force approaches have limited applicability. Luckily, there have been many advances in the root finding algorithms for waveguides\cite{Anemogiannis1992,Bakhtazad1997,Kwon2004,Rodriguez-Berral2005,Rodriguez-Berral2004,Smith1992}.

If the function $f(z)$\cite{footnote_6} is analytic, possesses no poles on and within the closed contour $C$ and finally if $f(z)$ does not go to zero on $C$ then:
\begin{equation}
\sum_{k=1}^N z_k^m = \frac{1}{2\pi\mi}\oint\nolimits_C z^m \: \frac{f'(z)}{f(z)} \: \dif z \label{contour}
\end{equation}
where $z_k$ denote the zeros of $f(z)$ in $C$, $N$ is the total number of zeros in $C$ and $m$ is an arbitrary nonnegative integer. Specifically, for $m=0$ one gets the total number of zeros within $C$. Knowing the number of zeros in a closed region enables one to dismiss regions of the complex search space in which there are no zeros. Furthermore, given the number of zeros in a region, one can do a subdivision until there is only one zero in the region of interest as described in Ref. \onlinecite{Kwon2004}. For those regions with a single zero, another contour integral as in \eqref{contour} with $m=1$ will give the location of the zero. If better numerical accuracy is desired, one can do a local search given the approximate position obtained by the contour integration. This method of finding the zeros of a function by repeated integration on the complex plane is called the \textsl{argument principle method} (\acro{APM}).

Note that for the \acro{APM} to work, the function $f(z)$ should be analytic in $C$. That requires no branch points to exist in the closed contour. Our implementation takes $\kappa_m$ as the variable of interest. Equations \eqref{dispersion} and \eqref{dispersionPEC} are rewritten in terms of only $\kappa_m$ by using \eqref{kvectors}. Singularities of $\tanh(\kappa_i a)$ and of  $\tanh(\kappa_m h)$ are removed by multiplying both sides of \eqref{dispersion} by $\cosh(\kappa_i a)$ and of \eqref{dispersionPEC} by $\cosh(\kappa_i a) \cosh(\kappa_m h)$.  At this stage the equations look like:
\begin{eqnarray}
	\epsilon_m \kappa_i \sinh(\kappa_i a) + \epsilon_i \kappa_m \cosh(\kappa_i a)= 0  \label{apm1}\\
	\begin{aligned}	
	\epsilon_m \kappa_i \sinh(\kappa_i a) \cosh(\kappa_m h) + \mspace{100mu} \\
	\epsilon_i \kappa_m \cosh(\kappa_i a)\sinh(\kappa_m h)=0  
	\end{aligned}\label{apm2}
\end{eqnarray}
with $\kappa_i = \sqrt{\kappa_m^2-k_0^2(\epsilon_m-\epsilon_i)}$ where $k_0=2\pi/\lambda$. The only branch points\cite{footnote_7} are those caused by the square root function in the definition of $\kappa_i$. Note that the function $\sqrt{z}\sinh(\sqrt{z})$ is single valued everywhere on the complex plane, since regardless of the choice for the sign of the square root, the result does not change. Similarly $\cosh(\sqrt{z})$ is single valued due to the evenness of the hyperbolic cosine function. Therefore, \eqref{apm1} and \eqref{apm2} have no branch points and are analytic in the whole complex plane.

In our implementation, we also did the conformal mapping $\kappa_m=\me^z$ to map the complex plane into strips. $z = \log(\kappa_m) \equiv \log\abs{\kappa_m} + \mi \arg(\kappa_m) = \log\abs{\kappa_m} + \mi \arg(\kappa_m) + \mi 2 \pi m$ for any integer $m$. The logarithm function is multi-valued, and repeats itself in strips that result from the $\mi 2 \pi m$ term. For instance, when we are interested in finding proper modes with $\Real(\kappa_m) >0$ we only need to search the strip $-\pi/2<\Imag(z)<\pi/2$.

Lastly, using computer algebra systems that can do symbolic mathematical manipulations, one can easily calculate $f^\prime(z)$ from the definition of $f(z)$ given by \eqref{apm1} or \eqref{apm2} in addition to doing numerical integrations on the complex plane.

\section{Some Details of the Mode Matching Algorithm}
Once the full set of modes are found using \acro{APM}, implementation of the mode-matching algorithm reduces to the calculation of the overlap integrals to build the matrix equation that should be solved to get the reflection and transmission coefficients, $R_{k p}$ and $T_{k p}$, of \eqref{junctionH}-\eqref{junctionE}. Since we have the analytical solutions for the fields as given in \eqref{field} and \eqref{fieldPEC}, overlap integrals $\Omega_{\{L,R\}}^{(m)}$ and  $[\field{e}{\{L,R\}}{m}|\field{h}{\{R,L\}}{k}]$ can be calculated analytically in closed form. The expressions are too long to reproduce here, but a computer algebra system can do the analytical manipulations. Once we have closed form results for the overlap integrals, formation of the matrix equation is very quick. The matrices have relatively small sizes and the solution of the linear matrix equation proceeds quickly. Also note that we numerically check the validity of the mode orthogonality condition of \eqref{eq17} before we start the mode-matching calculations so as to check the correctness of the \acro{APM} implementation.  We used Mathematica to implement \acro{APM} and the mode-matching method. The source code will be made available on the Internet under the \acro{GNU} general public license.

\bibliographystyle{apsrev}

\begin{thebibliography}{108}
\bibitem[{foo({\natexlab{a}})}]{footnote_1}
\bibinfo{note}{{Here, we are assuming that the fields are time harmonic with
  $\exp(+ \mi \omega t)$ time dependence. For $\exp(- \mi \omega t)$ time
  dependence, we get a positive imaginary value for the permittivity}}.
\bibitem[{foo({\natexlab{e}})}]{footnote_5}
\bibinfo{note}{{Here, we are assuming that we are operating at a frequency well
  above the plasma frequency, $\omega_p$, where $\epsilon_m = |\epsilon_m| < 1$
  so that the \acro{MIM} geometry describes a conventional dielectric slab
  waveguide as illustrated in the introduction section}}.
\bibitem[{foo({\natexlab{f}})}]{footnote_6}
\bibinfo{note}{{The variable $z$ in this appendix denotes an arbitrary complex
  number in the complex plane. It is \textsl{not} the space coordinate used in
  the rest of the paper}}.
\end{thebibliography}


\end{document}